\documentclass[10pt]{article}

\usepackage{amsmath}
\usepackage{graphicx}
\usepackage{cite}
\usepackage{namedplus}
\usepackage{color}

\usepackage{setspace}
\usepackage[labelfont=bf,labelsep=period,justification=raggedright]{caption}

\makeatletter
\renewcommand{\@biblabel}[1]{\quad#1.}
\makeatother

\date{}

\usepackage[amssymb,squaren]{SIunits}
\usepackage{xspace}
\usepackage{url}

\newcommand{\Ref}[1]{\ref{#1}}
\newcommand{\RefEq}[1]{Eq.~\Ref{#1}}
\newcommand{\RefEqs}[1]{Eqs.~\Ref{#1}}
\newcommand{\RefFig}[1]{Fig.~\Ref{#1}}
\newcommand{\RefFigs}[1]{Figs.~\Ref{#1}}

\newcommand{\ie}{i.e., }
\newcommand{\eg}{e.g., }
\newcommand{\avg}[1]{\left< #1 \right>} 

\definecolor{grey}{rgb}{0.5,0.5,0.5}

\begin{document}

\begin{flushleft}
{\Large \textbf{Emergence of Connectivity Motifs in Networks of Model \\Neurons with Short- and Long-term Plastic Synapses} }
\\
Eleni Vasilaki$^{1,2}$, Michele Giugliano$^{1,2,3,\ast}$,
\\
\bf{1} Dept. Computer Science, University of Sheffield, S1 4DP Sheffield, UK.\\
\bf{2} Theoretical Neurobiology and Neuroengineering Laboratory, Dept. Biomedical Sciences, University of Antwerp, B-2610 Wilrijk, Belgium.\\
\bf{3} Laboratory of Neural Microcircuitry, Brain Mind Institute, EPFL, CH-1015 Lausanne, Switzerland.\\
E-mail: $\ast$ michele.giugliano@ua.ac.be
\end{flushleft}

\section*{Abstract}
Recent evidence in rodent cerebral cortex and olfactory bulb suggests that short-term dynamics of excitatory synaptic transmission is correlated to the occurrence of stereotypical connectivity motifs. In particular, it was observed that  neurons with short-term facilitating synapses form predominantly reciprocal pairwise connections, while neurons with short-term depressing synapses form unidirectional pairwise connections. The cause of these structural differences in excitatory synaptic microcircuits is unknown.

We propose that these connectivity motifs emerge from the interactions between short-term synaptic dynamics (SD) and long-term spike-timing dependent plasticity (STDP). While the impact of STDP on SD was shown in simultaneous neuronal pair recordings {\em in vitro}, the mutual interactions between STDP and SD in large networks is still the subject of intense research. Our approach combines a SD phenomenological model with a STDP model that captures faithfully long-term plasticity dependence on both spike times and frequency.  As a proof of concept, we explore {\em in silico} the impact of SD-STDP in recurrent networks of spiking neurons with random initial connection efficacies and where synapses are either all short-term facilitating or all depressing. 
For identical background inputs, and as a direct consequence of internally generated activity,  we find that networks with depressing synapses evolve unidirectional connectivity motifs, while networks with facilitating synapses evolve reciprocal connectivity motifs. The same results hold for heterogeneous networks including both facilitating and depressing synapses.
Our study highlights the conditions under which SD-STDP explains the correlation between facilitation and reciprocal connectivity motifs, as well as between depression and unidirectional motifs. These conditions may lead to the design of experiments for the validation of the proposed mechanism.

\section*{Author Summary}
Understanding the brain requires unveiling its synaptic wiring diagram, which encodes memories and experiences. Synapses are,  however, more than mere plugs connecting neurons; they are communication channels implemented by biophysical systems that display history-dependent properties. During prolonged repeated activation, synapses may {\em functionally} undergo ``fatigue" or  may ``warm up". These effects, known as short-term plasticity, alter temporarily the communication properties of a synapse, \ie over tens to hundreds of milliseconds. At the same time, synapses undergo {\em structural} changes that persist from hours to days, known as long-term plasticity.

Recent progress in electrophysiology enabled simultaneous access to the synaptic wiring diagram in microcircuits, and to  its short-term features, \ie the ``fatigue" or ``warm-up" properties. Non-random connectivity patterns were reported, as i) co-occurrence of ``fatigue" features and unidirectional connections between two neurons, or ii) co-occurrence of ``warm-up" features and reciprocal connections.
We present a biologically plausible explanation for those patterns, based on the interaction of short- and long-term synaptic plasticity. We further formulate a computational model and provide an interpretation of the simulation results by means of its statistical description.

\section*{Introduction} \label{intro}

Among the most exciting challenges in Neuroscience, the investigation of the brain wiring diagram known as {\em connectomics} made spectacular progresses and generated excitement for its perspectives \cite{Seung2009}. Novel discoveries in molecular biology {\cite{Wickersham2007,Zhang2007,Lichtman2008}}, neuroanatomical methods, \cite{DenkHorstmann2004,Chklovskii2010}, electrophysiology \cite{Song2005,Hai2010,Perin2011}, and imaging \cite{Friston2011,Minderer2011,Wedeen2012} have pushed forward the technological limits, for an ultimate access to neuronal connectivity. This is in fact the most important level of organization of the brain \cite{Kandell2008}, pivotal to understanding the richness of its high-level cognitive, computational, and adaptive properties, as well as its dysfunctions.

At the microcircuit level  \cite{Silberberg2005,Grillner2005,DouglasMartin2007a,DouglasMartin2007b,Binzegger2004}, the non-random features of cortical connectivity have  recently raised a lot of interest \cite{Song2005,Perin2011}. The occurrence of stereotypical connectivity motifs was experimentally demonstrated and, in some cases, accompanied by physiological information on neuronal and synaptic properties \cite{Song2005,Wang2006,Silberberg:2007,Perin2011}, on activity-dependent short-term \cite{TsodyksMarkram:1997,Varela1997} and long-term plasticity \cite{BuonomanoMerzenich1998,MarkramGerstnerSjostrom2011} and rewiring \cite{Chklovskii2004,LeBe2006}.
These physiological details are complementary to anatomical connectivity mapping, and are known to underlie structural, dynamical, and computational network properties \cite{HaeuslerMaas2009}. In fact, information transmission between neurons in the brain takes place by means of more than mere ``connectors". For instance, short-term dynamics (SD) of synaptic efficacy is very common and has been quantified as transient and reversible facilitation or depression of postsynaptic responses, upon repeated presynaptic activation \cite{Cowan2003}. 
Relaying information to downstream neurons in a microcircuit is thus dependent on past-history, determined by presynaptic activation frequencies, and shaped by the SD profile at each synapse \cite{TsodyksMarkram:1997}.  

We address recent experimental findings obtained in rodent cortices \cite{Wang2006}, where short-term facilitation and depression were found to correlate to specific, pairwise, connectivity motifs: neurons, connected by synapses exhibiting short-term facilitation, form predominantly reciprocal motifs; neurons, connected by synapses exhibiting short-term depression, form unidirectional motifs. Interestingly, the same  over expression of connectivity motifs has been observed in another brain area, i.e. the excitatory microcircuitry of the olfactory bulb \cite{Pignatelli2009}. The cause of these structural differences are  largely unknown.

Inspired by the theory by Clopath {\em et al.} (2010) on the relationship between neural code and cortical connectivity \cite{Clopath2010}, we hypothesise that interactions between short-term and long-term synaptic plasticity could explain the observed pairwise connectivity motifs. In \cite{Clopath2010}, spike-timing dependent long-term synaptic plasticity (STDP) was shown to account {\em in silico} for the emergence of  non-random connectivity motifs in networks of model neurons  \cite{Clopath2010}. 
Adopting the same framework, we add to it a standard  phenomenological description of SD \cite{Tsodyks:1998} , and study SD-STDP interactions in recurrent networks of Integrate-and-Fire neurons \cite{BretteGerstner2005}.  

Extensive computer simulations of the evolution of synaptic efficacies under both external and self-generated activity have revealed an over-expression of reciprocal motifs on facilitating synapses, and of unidirectional motifs on depressing synapses. The departure from the initial random connection efficacies results from the interplay between neuronal activity and synaptic plasticity mechanisms, over long (STDP) and short time scales (SD). Supported by a mean-field analysis \cite{Tsodyks:2003,Renart2004,PfisterGerstner:2006,BarakTsodyks:2007,Gjorgjieva2011}, we conclude that the SD-STDP interplay is controlled by the distribution of firing rates in the network, through the switch of STDP from a correlational ``pre-post" {\em temporal} mode at low firing rates, to a ``Hebbian" {\em rate} mode at high firing rates, earlier reported experimentally by Sj\"{o}str\"{o}m {\em et al.} \cite{Sjostrom2001}. Short-term facilitation provides a positive feedback in recurrent microcircuits and it boosts firing, while depression acts as a negative feedback and it effectively attenuates high firing rates and discouraging their reverberations.
In the ``Hebbian" mode, STDP mostly gives rise to long-term potentiation (LTP) of synaptic efficacy, which becomes stronger resulting in reciprocal motifs. This leads to increasing firing rates of bidirectionally coupled neurons, with LTP acting as a long-term positive feedback, until the connection efficacies are ultimately stabilized by saturation. 
In the ``pre-post" (temporal) mode instead, it is the spike timing that drives LTP, which strongly favors the emergence of spatially asymmetric, \ie unidirectional, motifs \cite{Clopath2010}.

\section*{Materials and Methods} \label{Methods}
We study the properties of networks of excitatory spiking neurons \cite{BretteGerstner2005}, connected via plastic, {\em current-based}, synapses\cite{TsodyksMarkram:1997,Tsodyks:1998,BarakTsodyks:2007,PfisterGerstner:2006,Clopath2010,Gjorgjieva2011} via a set of simulations. We introduce a convention for distinguishing between ``strong" and ``weak" connections consistent with \cite{Clopath2010}, and a simple measure to quantify the occurrence of pair-wise motifs in our simulations. We finally provide a Wilson-Cowan firing rate model that is helpful for the interpretation of the numerical results. The numerical values of all parameters  are indicated in  Table \ref{tab:params}.

\subsubsection*{Neuron model} 
The network is composed of identical adaptive exponential Integrate-and-Fire neurons \cite{BretteGerstner2005}, each described by a membrane potential $V_m(t)$ and by a spike-frequency adaptation variable $x(t)$\cite{LiuWang2001}. Below a threshold $V_\theta$,  $V_m(t)$ satisfies the charge-balance equation
\begin{equation}
 c_m\ \dot{V}_m\ =\ g_{leak}\ \left(E_{leak}\ -\ V_m\right)\ +\ g_{leak}\ \Delta_T\ e^{\left(V_m\ -\ V_T\right)\ /\ \Delta_T}\ -\ x\ +\ I_{syn}\ +\ I_{ext}
 \label{chargebalance}
\end{equation}
\noindent  where $I_{syn}$ is the synaptic input from other neurons and  $I_{ext}$ the external (background) input currents. When a spike occurs, \ie $V_m(t)$ crosses $V_\theta$, $V_m$ is reset to a $E_{reset}$.

The spike-frequency adaptation variable $x(t)$ evolves as
\begin{equation}
 \tau_x\ \dot{x}\ =\ a\ \left(V_m\ -\ E_{leak}\right)\ -\ x
\label{adaptation}
\end{equation}
When a spike occurs, $x$ evolves as $x\ \rightarrow\ x\ +\ \Delta_x$. 
The numerical integration of \RefEq{chargebalance} is suspended for a period of time $\tau_{arp}$ following each spike, to mimic absolute refractoriness, during which $V_m$ remains ``clamped" at $E_{reset}$.

The model details are not essential to our conclusions (see Supplemental Information, Figure S7).

\subsubsection*{External (background) inputs} 
Each neuron is identified by an index $i=1,2,3\ldots$ and it receives a time variant input $I_{ext\ i}$ according to the following protocols.

{\bf{\emph{Toy Network:}}} ( \eg \RefFig{fig:ToyNetwork}B, 10 neurons) $I_{ext\ i}$ consists of a $0.5\ nA$ constant current, as well as periodic $1\ nA$ square pulses.  Pulses occur as in a traveling wave of activity, which moves every $5\ msec$ from one unit, \eg the $i$-th neuron, to its next index neighbour, \eg the $(i+1)$-th neuron, see also \cite{Clopath2010}. Each pulse is delivered in turn to all neuronal indexes, as an extremely narrow bell-shaped profile with unitary amplitude and standard-deviation of $0.5$, resulting in neighbouring neurons being only slightly stimulated. Each pulse is of sufficient amplitude to elicit neuronal firing in the unit where the bell-shaped profile is centred on, \eg the $i$-th unit (unless refractory).

 {\bf{\emph{Large Network:}}} (\eg  \RefFig{fig:PopulationsMicroscopic}, 1000 neurons) $I_{ext}$ is as in the toy network with the addition of an uncorrelated gaussian noisy current \cite{AmitBrunel1997}, with mean $\mu$, standard deviation $\sigma\ =\ 200\ pA$, and autocorrelation time length $\tau_{syn}=5\ msec$. Parameter $\mu$ is drawn randomly before launching the simulation and for each neuron of the network, from a gaussian distribution with mean $200\ pA$ and unitary coefficient of variation. The noisy current mimics asynchronous synaptic inputs from (not explicitly modelled) background populations \cite{Roxin2011,Tuckwell:1989}.

\subsubsection*{Internal (synaptic) inputs} 
Neurons connect to each other according to a fixed anatomical wiring matrix $[C_{ij}]$, which indicates whether neuron $j$-th projects to neuron $i$-th, \ie $C_{ij}\ =\ 1$, or not, \ie $C_{ij}\ =\ 0$. The matrix $[C_{ij}]$ is obtained from an all-to-all connectivity without autapses (\ie $C_{ii}\ =\ 0$), upon randomly pruning $\approx 20\%$ of its elements (see e.g. \RefFig{fig:ToyNetwork}B), to introduce variability in neuronal firing. A more substantial reduction of the structural connectivity does not affect qualitatively our conclusion, although it downscales the number of  plastic synapses available for statistical analysis. This would require larger networks to be simulated in order to obtain the same statistical confidence over our results,  and would require upscaling $G_{ij}$ (\eg through increasing $W_{max}$, see the next subsections) to obtain the same network firing rates.

The $i$-th neuron receives at any time a synaptic current $I_{syn\ i}$, described as
\begin{equation}
 \dot{I}_{syn\ i}\ =\ -I_{syn\ i}\ /\ \tau_{syn}\ +\ \sum_{j=1}^N \sum_{f}^\infty\ C_{ij}\ G_{ij}\ \delta(t-t^{f}_{j})
\label{synaptic_coupling}
\end{equation}

\noindent where $t^{f}_{j}$  represents the occurrence time of the $f$-th spike emitted by the $j$-th presynaptic neuron, and $G_{ij}$ the amplitude of the postsynaptic current (PSC), corresponding to the activation of the synapse by the presynaptic  $j$-th neuron. The Dirac's delta function $\delta(t)$ represents the occurrence of a presynaptic action potential. 
\RefEq{synaptic_coupling} models incoming individual PSCs as waveforms with instantaneous rise time and slower decay time \cite{Sterrat2011}, imposes a linear postsynaptic superposition of effects, and implies that in the lack of any presynaptic spiking activity, $I_{syn\ i}$ decays exponentially to zero with a time-constant $\tau_{syn}$, and when a presynaptic spike is fired it evolves as $I_{syn\ i}\ \rightarrow\ I_{syn\ i}\ +\ C_{ij}\ G_{ij}$.

\subsubsection*{Frequency-dependent short-term synaptic dynamics (SD)} 
$G_{ij}$ defines the amplitude of the PSC from presynaptic neuron $j$-th to postsynaptic neuron $i$-th. On short timescales, its value changes as a function of the activation history of the presynaptic neuron, leading to transient and reversible depression or facilitation of synaptic efficacy \cite{TsodyksMarkram:1997}. This is referred here as homosynaptic plasticity and implemented by having $G_{ij}$ proportional to the amount of used resources for neurotransmission $u_{ij}\ r_{ij}$ and their maximal availability $A_{ij}$, \ie $G_{ij}\ =\ A_{ij}\ u_{ij}\ r_{ij}$.

Frequency-dependent short-term synaptic dynamics is described by the following equations, with a different set of parameter values to capture depressing or facilitating synapses \cite{TsodyksMarkram:1997}:
\begin{equation}
  \left\{
    \begin{array}{l}
     \dot{r}_{ij}\ =\ (1-r_{ij}) / \tau_{rec}\ -\ \sum_{k_j}^\infty\ u_{ij}\ r_{ij}\ \delta(t-t_{k_j})\\[3mm]
     \dot{u}_{ij}\ =\ -u_{ij} / \tau_{facil}\ +\ \sum_{k_j}^\infty\ U\ (1-u_{ij})\ \delta(t-t_{k_j})\\[3mm]
      \end{array}
      \right.
 \label{STPrule}
\end{equation}

For the sake of notation, indexes have been dropped from parameter $U$ as well as from the time constants $\tau_{rec}$ and $\tau_{facil}$ in \RefEqs{STPrule}, although in general each synapse has its own parameters (see Table \ref{tab:params}). \RefEqs{STPrule} reduce to the following update rules: i) when no spike is fired by the presynaptic neuron $j$, $u_{ij}$ and $r_{ij}$ recover exponentially to their resting values, $U$ and $1$, respectively; ii) as a presynaptic spike occurs, $r_{ij}$ is reduced as $r_{ij}\ \rightarrow (1-u_{ij})\ r_{ij}$, while $u_{ij}$ is increased as $u_{ij}\ \rightarrow u_{ij}\ +\ U_{ij}\ (1-u_{ij})$. 
The impact of short-term plasticity of PSCs amplitude is exemplified in \RefFig{fig:ToyNetwork}C,F. 
\subsubsection*{Spike-timing dependent long-term plasticity (STDP)} 
We further extend the description of PSCs (\RefEqs{synaptic_coupling}-\ref{STPrule}), by an additional scaling factor $W_{ij}$, to incorporate STDP \cite{MarkramGerstnerSjostrom2011}, see also \cite{Morrison2008}: 
\begin{equation}
G_{ij}\ =\ W_{ij}\ A_{ij}\ u_{ij}\ r_{ij}. 
\end{equation}
$W_{ij}$ changes on timescales longer than $\tau_{rec}$ and $\tau_{facil}$ and according to the correlated activity of both pre- and postsynaptic neurons, as in \cite{PfisterGerstner:2006}, capturing spike-triplets effects \cite{PfisterGerstner:2006,Clopath2010,Gjorgjieva2011}. This is referred here as heterosynaptic plasticity. Each neuron is complemented by four variables, \ie $q_1$, $q_2$, $o_1$, $o_2$, which act as running estimates of the neuron firing rate, averaged over distinct time scales, \ie $\tau_{q_1}$, $\tau_{q_2}$, $\tau_{o_1}$, $\tau_{o_2}$. In the lack of any firing activity of the $j$-th neuron, those variables exponentially relax to zero:
\begin{equation}
     \tau_{q_1}\ \dot{q}_{1_j}\ =\ -q_{1_j}\ \ \ \ \ \ \ \ \ \ 
     \tau_{q_2}\ \dot{q}_{2_j}\ =\ -q_{2_j}\ \ \ \ \ \ \ \ \ \ 
     \tau_{o_1}\ \dot{o}_{1_j}\ =\ -o_{1_j}\ \ \ \ \ \ \ \ \ \ 
     \tau_{o_2}\ \dot{o}_{2_j}\ =\ -o_{2_j}
\label{STDPindicators}
\end{equation}

\noindent while each time the neuron fires, these variables are increased by a unit: 
\begin{equation}
 q_{1_j}\ \rightarrow\ q_{1_j}\ +\ 1\ \ \ \ \ \
 q_{2_j}\ \rightarrow\ q_{2_j}\ +\ 1\ \ \ \ \ \
 o_{1_j}\ \rightarrow\ o_{1_j}\ +\ 1\ \ \ \ \ \
 o_{2_j}\ \rightarrow\ o_{2_j}\ +\ 1. 
\label{STDPindicators2}
\end{equation}

\noindent This enables a compact formulation of STDP: when the $j$-th neuron spikes, the following updates are performed for all the indexes $i$:
\begin{equation}
  \left\{
    \begin{array}{l}
     W_{ij}\ \rightarrow\ W_{ij}\ -\eta\ \ o_{1_i}(t)\ \left[ A_2^-\ +\ A_3^-\ q_{2_j}(t-\epsilon)\right]\\
     W_{ji}\ \rightarrow\ W_{ji}\ +\eta\ \ q_{1_i}(t)\ \left[ A_2^+\ +\ A_3^+\ o_{2_j}(t-\epsilon)\right]\\
      \end{array}
      \right.
 \label{STDPrule}
\end{equation}

\noindent as the $j$-th neuron is presynaptic to all connected  $i$-th neurons, and postsynaptic to all connected $i$-th neurons, respectively. Numerically, the evaluation of $q_{2_j}$ and $o_{2_j}$ is performed just before the neuron $j$-th spikes, as indicated by the infinitesimal time-advance notation of $\epsilon$. When no spike occurs, $W_{ij}$ maintains indefinitely its value. The instantaneous value of each $W_{ij}$ is bounded in the range $[0\ ;\ W_{max}]$. 
Unless otherwise stated, $W_{ij}$ is initialized from a uniform random distribution between $0$ and $W_{max}$, prior to the start of each simulation.

We remark that our choice for combining STDP and SD together aims at a phenomenological description of their interactions. It is assumed that all STDP variables are local as well as that STDP and SD models act independently. This allows us to easily analyse network interactions and avoid the paradox of pre-synaptic firing detection under strong short-term depression of synaptic efficacy. In fact, we assume for simplicity that the STDP coincidence-detector is located at each synapse and that such a detector is able to access timing information of presynaptic spikes, regardless on whether they result in large or small PSCs.  Similarly, the same coincidence-detector will also sense all the spikes emitted by the post-synaptic neuron. This is the only information required to calculate the long-lasting synaptic modification.

\subsubsection*{Convention on ``strong" and ``weak" connections and network symmetry index} 

In this study, we focus on the appearance or disappearance of a strong connection between two neurons, only for units that are anatomically connected, \ie $C_{ij}\ =\ 1$. For the sake of comparison, we adopted the framework of Clopath {\em et al.} (2010)\nocite{Clopath2010}, where the activity-dependent appearance or disappearance of a connection conventionally occurs in terms of a competition among the ``strong" links in a ``sea" of weak synapses \cite{Song2005}. As in their paper, we adopt the convention of identifying as ``strong" those connections whose corresponding STDP synaptic factor $W_{ij}$ is above the $2/3$ of its upper bound $W_{max}$.

With such a definition, we measure the symmetry in the network connectivity by counting reciprocal or unidirectional motifs as (but see \eg \cite{HubertBaker:1979} for alternative definitions)
\begin{equation}
	s(W)\ =\ 1\ -\ \left(0.5\ N \left(N-1\right)\ -\ M\right)^{-1}\ \sum_{i=1}^{N}\sum_{j=i+1}^{N}{|W^*_{ij}\ -\ W^*_{ji}|}
\label{symmetry_measure}
\end{equation}

\noindent where $N$ is the size of the matrix $W$, as well as the size of the network.
The symmetry index $s(W)$ takes values in the range $[0\ ;\ 1]$ and depends on the average {\em similarity} between elements of $W$ that are on symmetric positions with respect to the diagonal. Following our previous convention, $W_{ij}$ and $W_{ji}$ are first normalized and then zero-clipped: $W^*_{ij}\ =\ W_{ij}\ /\ W_{max}$ if $W_{ij}\ >\ 2/3\ W_{max}$, and otherwise $W^*_{ij}\ =\ 0$. In \RefEq{symmetry_measure}, $M$ represents the number of null pairs $\{W^*_{ij},\ W^*_{ji}\}\ =\ \{0,\ 0\}$ that occur as a consequence of clipping or by initialization.
Evaluating $s(W)$ on networks with a majority of unidirectional connections results in values close to $0$ (\eg \RefFig{fig:ToyStatistics}A), while its evaluation on networks with a majority of reciprocal connections results in values close to $1$ (\eg \RefFig{fig:ToyStatistics}B). For uniform random matrices $W$, the full statistics of $s(W)$ can be calculated (see Supplemental Information) and employed for deriving a significance measure for $s$ as a $p$-value, representing the probability that the value of $s$ observed in simulations could result by chance.
This symmetry measure is useful only {\em in silico}, where the entire connectivity matrix as well as its maximal synaptic efficacy are known. The same measure cannot be considered as an operational tool for quantifying experimentally-extracted microcircuit connectivity, where statistical counting of connectivity motifs and their comparison to chance-level are relevant \cite{Song2005}. Approaching a qualitative comparison to the experiments of Wang {\em et al.} (2006)\nocite{Wang2006}, such a statistical counting has been used in the large simulations of \RefFig{fig:PopulationsMicroscopic}.

\subsubsection*{Mean-field Network Description}
We approximate the firing rate of the network of Integrate-and-Fire units through its mean-field dynamical description \cite{WilsonCowan:1972,DayanAbbott:2001,Renart2004}, closely following earlier work \cite{Tsodyks:1998,BarakTsodyks:2007}. We consider the simplifying hypotheses that i) the network consists of one or more non overlapping subpopulations of excitatory neurons (see \RefFigs{fig:MeanField}A-B and \RefFig{fig:PopulationsMacroscopic}A) and that ii) neurons within each subpopulation share identical synaptic coupling, connectivity, and short-term synaptic plasticity properties, \ie all depressing or all facilitating, as in \RefFig{fig:PopulationsMacroscopic}A; see \RefFig{fig:MixedMeanField}A for an exception. We also assume that iii) for each of the (sub)populations, the individual neuronal firing follows a Poisson distribution, with instantaneous mean firing rate $E(h)$, which depends monotonically on the corresponding overall input currents $h$. Under these hypotheses, neurons can be only distinguished by the subpopulation they belong to, \ie depressing ÒDÓ or facilitating ÒFÓ, and their firing can be identified as $E_D(h_D)$ and $E_F(h_F)$. For the case of two subpopulations (\RefFig{fig:PopulationsMacroscopic}A), $h_D$ and $h_F$ evolve over a characteristic time scale $\tau$ as
\begin{equation}
  \left\{
    \begin{array}{l}
     \tau\ \dot{h}_D\ =\ -h_D\ +\ J_{DD}\ E_D\ +\ J_{DF}\ E_F\ +\ \hat{I}_{ext}\\ [3mm]
     \tau\ \dot{h}_F\ =\ -h_F\ +\ J_{FD}\ E_D\ +\ J_{FF}\ E_F\ +\ \hat{I}_{ext}\\ [3mm]
    \end{array}
  \right.
\label{mean_field_dyn}
\end{equation}

\noindent where $\hat{I}_{ext}$ represents the external input, $J_{DD}$ and $J_{FF}$ the average synaptic efficacies of recurrent connections within each subpopulation, and $J_{DF}$ and $J_{FD}$ the average synaptic efficacies of inter-subpopulations connections. On a first approximation, $J$ can be considered as the ensemble average of $G_{ij}$. 
Firing rates $E_D$ and $E_F$ are computed from $h_D$ and $h_F$ as threshold-linear frequency-current response functions: $E_D\ =\ \left[h_D - \theta\right]_+$ and $E_F\ =\ \left[h_F - \theta\right]_+$, with $\left[x\right]_+= max\{x; 0\}$ (for alternatives see \cite{LaCamera2008,Giugliano2008}). 
$J_{DD}$, $J_{FF}$, $J_{DF}$, $J_{FD}$ undergo plastic changes, on both the short and long time scales: indicating by $J_{ab}$ the coupling between the presynaptic population $b$ and the postsynaptic population $a$, then $J_{ab}\ =\ \hat{W}_{ab}\ \hat{A}_{ab}\ \hat{u}_b\ \hat{x}_b$ with $a,b \in \{D,F\}$. 
The mean-field variables $\hat{u}_b$ and $\hat{x}_b$ depend only on the presynaptic firing rate $E_b$, and capture the short-term homosynaptic plasticity in the mean-field approximation of \RefEqs{STPrule} \cite{Tsodyks:1998}, as:
\begin{equation}
  \left\{
    \begin{array}{l}
     \dot{\hat{u}}_D\ =\ \left(U_D\ -\ \hat{u}_D\right)\ /\tau_{facil\ D}\ +\ U_D\ \left( 1\ -\ \hat{u}_D\right)\ E_D\\[3mm]
     \dot{\hat{x}}_D\ =\ \left(1\ -\ \hat{x}_D\right) / \tau_{rec\ D}\ +\ \hat{u}_D\ \hat{x}_D\ E_D\\[3mm]
     \dot{\hat{u}}_F\ =\ \left(U_F\ -\ \hat{u}_F\right) / \tau_{facil\ F}\ +\ U_F\ \left( 1\ -\ \hat{u}_D\right)\ E_D\\[3mm]
     \dot{\hat{x}}_F\ =\ \left(1\ -\ \hat{x}_F\right) /  \tau_{rec\ F}\ +\ \hat{u}_F\ \hat{x}_F\ E_F\\[3mm]
    \end{array}
  \right.
\label{mean_field_STP}
\end{equation}

On longer timescales, the mean-field approximation of STDP given in \cite{PfisterGerstner:2006} is adopted here by altering the factor $\hat{W}_{ab}$, which evolves as a function of both presynaptic $E_{b}$ and postsynaptic $E_{a}$ firing rates:
\begin{equation}
          \frac{1}{\eta}\ \dot{\hat{W}}_{ab}\ =\ -A_2^-\ \tau_{o_1}\ E_b\ E_a\ -\ A_3^-\ \tau_{o_1}\ \tau_{q_2}\ E_b^2\ E_a\ +\ A_2^+\ \tau_{q_1}\ E_b\ E_a\ +\ A_3^+\ \tau_{q_1}\ \tau_{o_2}\ E_b\ E_a^2\ .
\label{mean_field_STDP}
\end{equation}

\section*{Results}

We hypothesize that synaptic connectivity reflects the interaction of short-term plasticity, such as homosynaptic depression and facilitation \cite{TsodyksMarkram:1997}, with long-term plasticity. Inspired by a recent theory, which links connectivity to spiking activity and to the neural code \cite{Clopath2010}, we ask the  question: could known relationships between short-term plasticity and spiking activity in recurrent networks \cite{Tsodyks:1998,Tsodyks:2003,BarakTsodyks:2007} explain the occurrence of connectivity motifs  \cite{Wang2006,Pignatelli2009} observed experimentally? In support to our hypothesis, we present simulations of networks of excitatory  neuron models with activity-dependent plastic connections. We then study the necessary conditions for the emergence of these specific connectivity motifs by  standard firing rate models. 

\subsubsection*{A toy microcircuit model}

\begin{figure}[htb!]
\centering
\includegraphics[scale=0.65]{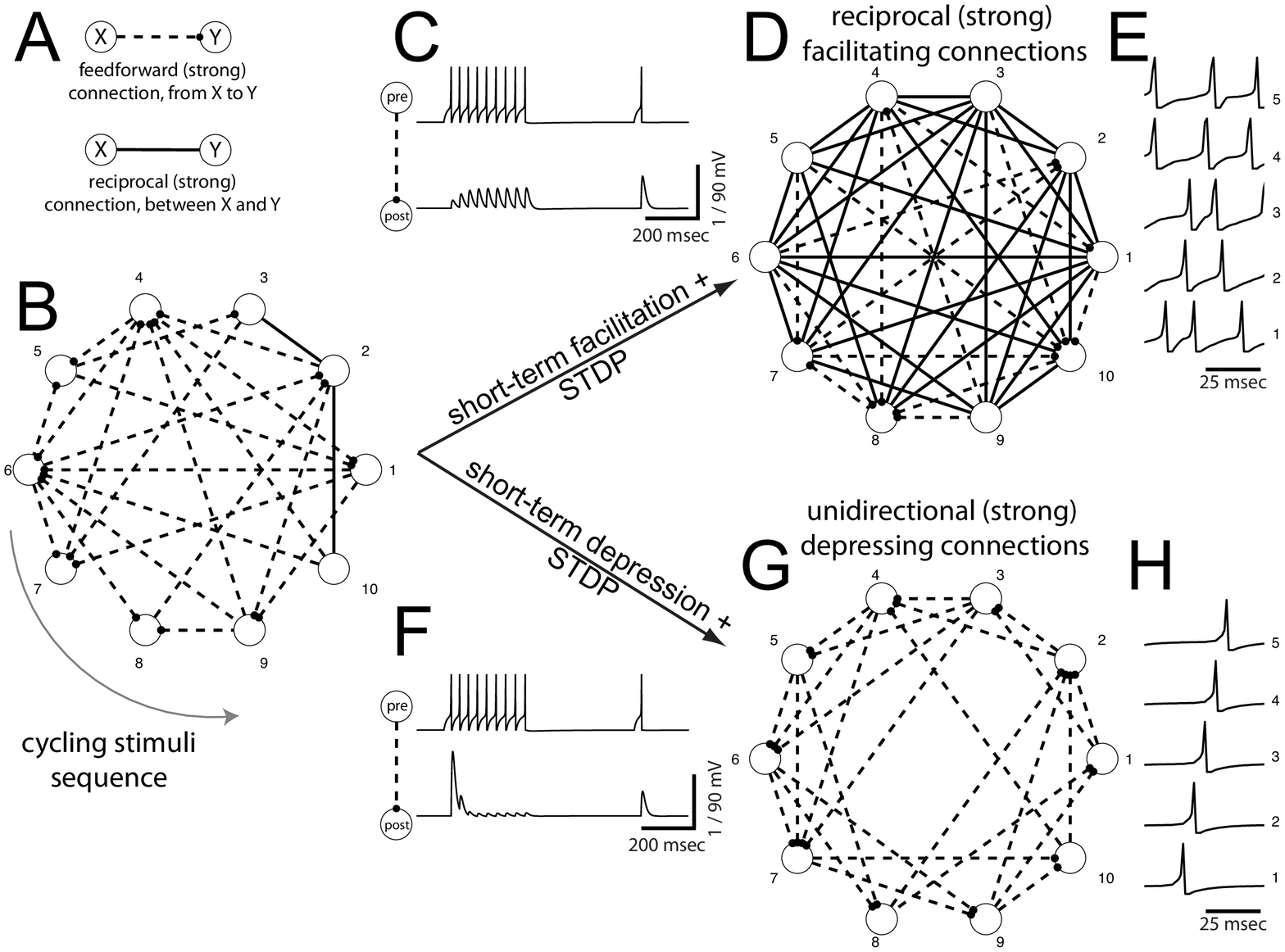}
\caption{{\bf Emergence of connectivity motifs in a toy model network}. Unidirectional (reciprocal) strong excitatory connections are indicated ({\bf A}) as dashed (continuous) line segments, representing the topology of the network ({\bf B}). Each model neuron receives periodic spatially alternating depolarizing current pulses, strong enough to make it fire a single action potential. Synapses among connected neurons display ({\bf C}) short-term facilitation of postsynaptic potential amplitudes. Spike-Timing Dependent Plasticity (STDP) leads to strengthened connections and result into a largely reciprocal topology ({\bf D}). Modifying the short-term plasticity profile into depressing ({\bf F}), leads to a largely unidirectional topology shaped by STDP({\bf G}). Distinct motifs of strong connections arise from short- and long-term plasticities, due to distinct firing patterns (compare {\bf E} and {\bf H}), under identical external stimulation and initial connections. Parameters: $A_{ij}=400pA$. 
\label{fig:ToyNetwork}}
\end{figure}

In order to demonstrate and test the key features of our hypothesis, we consider a simplified representation of excitatory neuronal microcircuits, as a network of adaptive exponential Integrate-and-Fire units \cite{BretteGerstner2005}. The validity of our results does not depend on the specific details of the model neuron and of its spike-frequency adaptation parameters chosen here (Supplemental Information, Figure S7).  Neurons are connected to each other through excitatory synapses (\RefFig{fig:ToyNetwork}B), whose efficacy undergoes either short- or long-term plasticity. Using a standard model for short-term synaptic dynamics (SD) \cite{TsodyksMarkram:1997}, we simulate homosynaptic, use-dependent modifications of the postsynaptic potentials (PSPs) amplitude. We also employ a recently introduced phenomenological description of heterosynaptic long-lasting potentiation or depression of PSPs, able to capture with great accuracy both spike-timing and frequency effects \cite{PfisterGerstner:2006}. We refer to this long-term associative plasticity as spike-timing dependent  (STDP). Both SD and STDP occur simultaneously with neuronal dynamics, although across distinct timescales. We note that STDP interacts with short-term plasticity as a frequency-independent scaling factor of PSPs amplitude, rather than contributing to a redistribution of synaptic efficacy in the sense of \cite{MarkramTsodyks1996,AbbottNelson2000} (see Discussion). In addition to excitatory PSPs evoked by trains of presynaptic spikes, neurons receive an external current input, deterministically played back over and over, as a traveling wave of activity (\RefFig{fig:ToyNetwork}B).  Such an external stimulation recreates the {\em temporal coding} of Clopath {\em et al.} (2010)\nocite{Clopath2010}, which imposes deterministic spike-timing correlations among evoked spikes (see Methods). The external input can be regarded an oversimplified generic \eg thalamic, input with known temporally correlated structure. 

We define two microcircuits, identical for all aspects of neuronal properties, maximal synaptic efficacy, anatomical connectivity, and external inputs, with the exception of the SD properties. Specifically, one microcircuit includes exclusively short-term facilitating synapses (\RefFig{fig:ToyNetwork}C), while the other includes only depressing synapses (\RefFig{fig:ToyNetwork}F). In order to describe changes in the microcircuit connectivity, we adopt the same framework of Clopath {\em et al.} (2010)\nocite{Clopath2010} where connections form or disappear via competition among the ``strong" links in a ``sea" of weak synapses \cite{Song2005} (see Methods). The connectivity, which is randomly initialized, slowly evolves into largely non-random configurations during the simulation (\RefFig{fig:ToyNetwork}D,G). At the steady-state, these configurations match the experimentally observed correlations: reciprocal motifs emerge in cell pairs more often than unidirectional motifs,  when synapses are facilitating; the opposite occurs when synapses are depressing. 
This is revealed both by direct inspection of the synaptic efficacy matrix $[W_{ij}]$ (\eg see \RefFig{fig:asinClopath}) and by quantification of its symmetry index $s$ (see Methods). Although other statistical measures can be defined, as for instance the percentage of weak elements in $[W_{ij}]$ or the sparseness of the emerging connectivity, here we focus on the reciprocal {\em versus} unidirectional features of its strong elements, according to the same convention of Clopath {\em et al.} (2010).
When $s$ takes values close to $1$, almost all of the existing pairwise anatomical connections are reciprocal and $[W_{ij}]$ is almost a symmetric matrix. On the other hand, for values of $s$ close to $0$, unidirectional or very weak connection motifs prevail. 

We underline that the prevalence of external inputs over recurrent synaptic inputs (or vice versa) is not imposed a priori but it depends on the interaction between SD and STDP. For facilitating microcircuits, the overall recurrent (synaptic) input that a neuron receives becomes progressively larger than the external input, as soon as more and more reciprocal connectivity motifs are established by STDP. For depressing microcircuits, the opposite holds, as more and more connections are weakened by STDP.

These results, summarized in \RefFig{fig:ToyNetwork}  for a sample microcircuit composed by ten neurons, have been confirmed over $2000$ repeated simulations, analysed in \RefFig{fig:ToyStatistics}. Over a slow timescale, STDP results in a very small degree of symmetry ($s\ =\ 0.01\ \pm\ 0.01$, mean $\pm$ stdev), with high significance ($p\ <\ 10^{-4}$) for each of the repeated simulations involving depressing synaptic connections (\RefFig{fig:ToyStatistics}C). Under identical external inputs to the microcircuit, STDP leads instead to a large degree of symmetry ($s\ =\ 0.61\ \pm\ 0.10$, mean $\pm$ stdev), with high significance ($p\ <\ 10^{-4}$) in about $75\%$ of the simulations involving facilitating synaptic connections (\RefFig{fig:ToyStatistics}D). In the remaining $25\%$ of the cases, the resulting symmetry value was in the range $[0.25 ; 0.55]$, suggesting that to some extent, the variability observed in experiments \cite{Wang2006,Pignatelli2009} was replicated in the simulations, where not $100\%$ of the times the motifs correlations emerged. In this simplified example, the variability is attributed mostly to the sparse anatomical connectivity and the irregular firing regimes this imposes. Doubling the simulation time (not shown) led to a very minor reduction of the variability on $s$, by less than $3\%$ and only for the facilitating synaptic connections, suggesting that stationarity had been already reached.

\subsubsection*{A simple mechanism for the emergence of motifs}

\begin{figure}[htb!]
\centering
\includegraphics[scale=0.6]{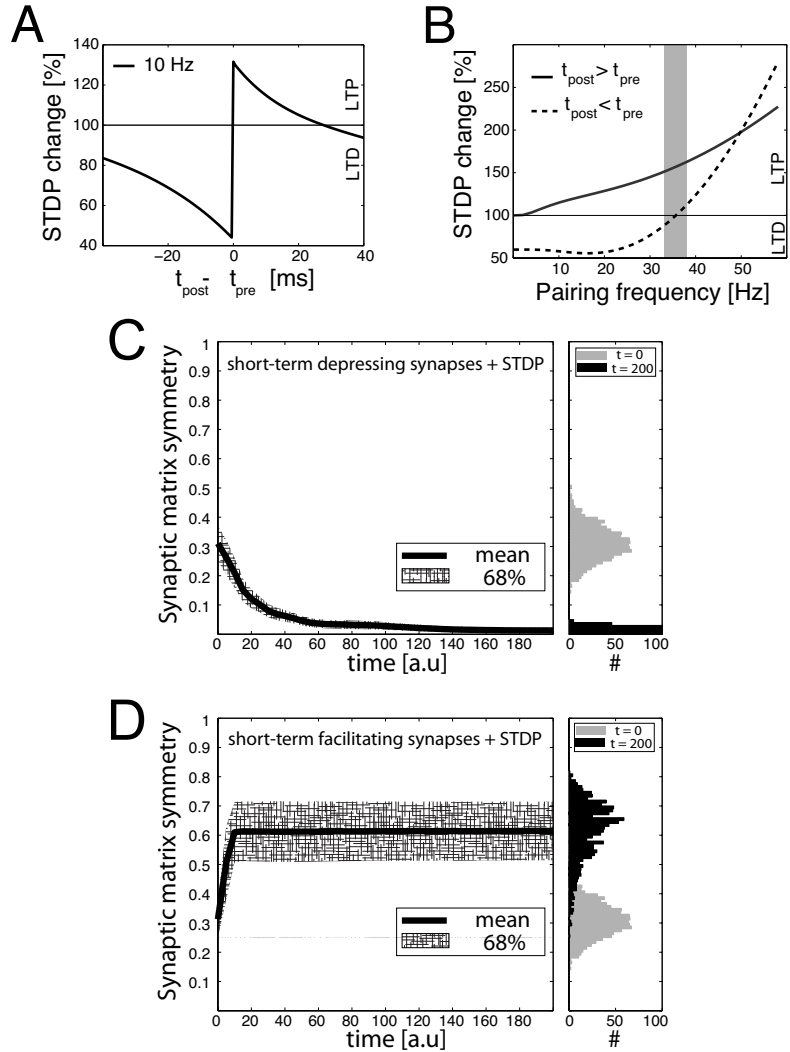}
\caption{{\bf Statistics of motifs emergence in a toy model network.}  
When decoupled from recurrent interactions, an isolated model synapse undergoes long-term changes depending on pre- and postsynaptic spike timing ({\bf A}) and pairing frequency ({\bf B}). Above a {\em critical} frequency (gray shading), spike timing no longer matters and long-term potentiation of synaptic efficacy (LTP) prevails on long-term depression (LTD). 
Panels {\bf C, D:} The simulations of \RefFig{fig:ToyNetwork} were repeated $2000$ times, each time starting from a random initial topology. STDP progressively induced a persisting non-random reconfiguration of strong connections, quantified across time by a symmetry index (see Methods). Neurons connected only by short-term depressing synapses evolved strong unidirectional connections, corresponding to a low symmetry index. This is displayed in panel {\bf C}, as an average across the $2000$ simulations (left panel). Initial and final distributions of symmetry index values are also shown (right panel, gray and black histogram respectively). Neurons connected only by short-term facilitating synapses evolved instead strong bidirectional connections with high  symmetry indexes ({\bf D}).
\label{fig:ToyStatistics}}
\end{figure}

In the microcircuits considered above, the statistics of the anatomical connectivity, the external inputs, as well as the single-neuron properties are identical. Any difference in long-term modification of connections could only arise i) from differences in the collective neuronal activity and take place ii) through the spike-timing and firing-rate dependent mechanisms underlying connection motifs formation, as first described in \cite{Clopath2010}. In the microcircuits considered here, neuronal activity is known to depend on the connectivity and the short-term changes in PSPs amplitude \cite{Tsodyks:1998}.  Indeed, during the $2000$ simulations of \RefFig{fig:ToyStatistics}, the microcircuits including short-term depressing synapses collectively fired at $20\ \pm\ 0\ Hz$ (see also \RefFig{fig:ToyNetwork}H), lower than the recurrent microcircuits employing short-term facilitating synapses, which  fired at $59\ \pm\ 4\ Hz$ (see also \RefFig{fig:ToyNetwork}E). 
Facilitating synapses  charge up with ongoing presynaptic activity and they recruit more connected neurons, as in a positive feedback. On the contrary, depressing synapses get soon fatigued, resulting into a weaker subsequent recruitment of postsynaptic neurons, whose firing would further depress synaptic efficacy, as in a negative feedback.

We identify such an asymmetry as the cause underlying motifs emergence, and the firing-rate dependence of STDP \cite{MarkramGerstnerSjostrom2011} as its mechanism. In particular, it is the functional switch of STDP from a correlational ``pre-post" {\em temporal} mode at low firing rates, to a ``Hebbian" {\em rate} mode at high firing rates, that entirely accounts for the asymmetry in the connectivity motifs.
In addition to the spike-timing information (\RefFig{fig:ToyStatistics}A), the STDP model employed here faithfully captures the strong dependency on the neuronal firing rates \cite{Sjostrom2001} (\RefFig{fig:ToyStatistics}B). 
For the sake of illustration, we isolate the impact on synaptic efficacy of \RefEqs{STDPindicators},\ref{STDPindicators2},\ref{STDPrule}, in a two-neuron system, with one neuron projecting to the other via a single synapse. We simulate the long-term change in PSPs amplitude at that synapse (see \RefFig{fig:ToyStatistics}B and \cite{PfisterGerstner:2006}), imposing $75$ pre-post spike pairing events, evoked at different uniform firing frequencies. We study two cases: (i) each presynaptic spike precedes the postsynaptic spike by $10\ msec$, \ie $t_{pre}<t_{post}$, or  (ii) vice versa, \ie $t_{pre}>t_{post}$.
In agreement with the experiments \cite{Sjostrom2001}, above $30-40\ Hz$ long-term potentiation (LTP) prevails on long-term depression (LTD), even in those cases when spike-timing {\em per se} would promote LTD, \ie $t_{pre}>t_{post}$. Below that {\em critical} frequency, LTP or LTD reflects causal or anti-causal relationships between pre- and post-synaptic firing times, respectively \cite{MarkramGerstnerSjostrom2011}. 

It follows that facilitating microcircuits would preferentially develop reciprocal connectivity motifs, irrespectively of the spike-timing, due to their higher firing rates that promote LTP. The consequent increased recurrence in the connections would further increase the overall population firing rate, above the {\em critical} frequency. This increment in the firing rates leads at first place to LTP. 
Depressing microcircuits instead do not develop preferentially reciprocal motifs, but reflect the asymmetric temporal structure of the input. On a first approximation, their connections would rather not promote persistent high activity regimes; instead they mostly respond to external stimuli. These stimuli are imposed here at low rates and give rise to feed-forward links, due to the repeating  traveling-wave features of the stimulation, as in \cite{Clopath2010}. 

These specific connectivity and activity configurations are stable when imposing reflecting boundaries, as in the numerical implementation of STDP \cite{Song2000,AbbottNelson2000,Rubin2001}. These boundaries limit the synaptic efficacy to a minimal and a maximal values (see Methods). We also remark that the firing rate dependence of STDP, shown in Fig. \ref{fig:ToyStatistics}B, is not captured by all the models proposed in the literature. In our case, the minimal set of STDP model features sufficient for motifs emergence must include the reversal of LTD into LTP at high firing rates.

\begin{figure}[htb!]
\centering
\includegraphics[scale=0.62]{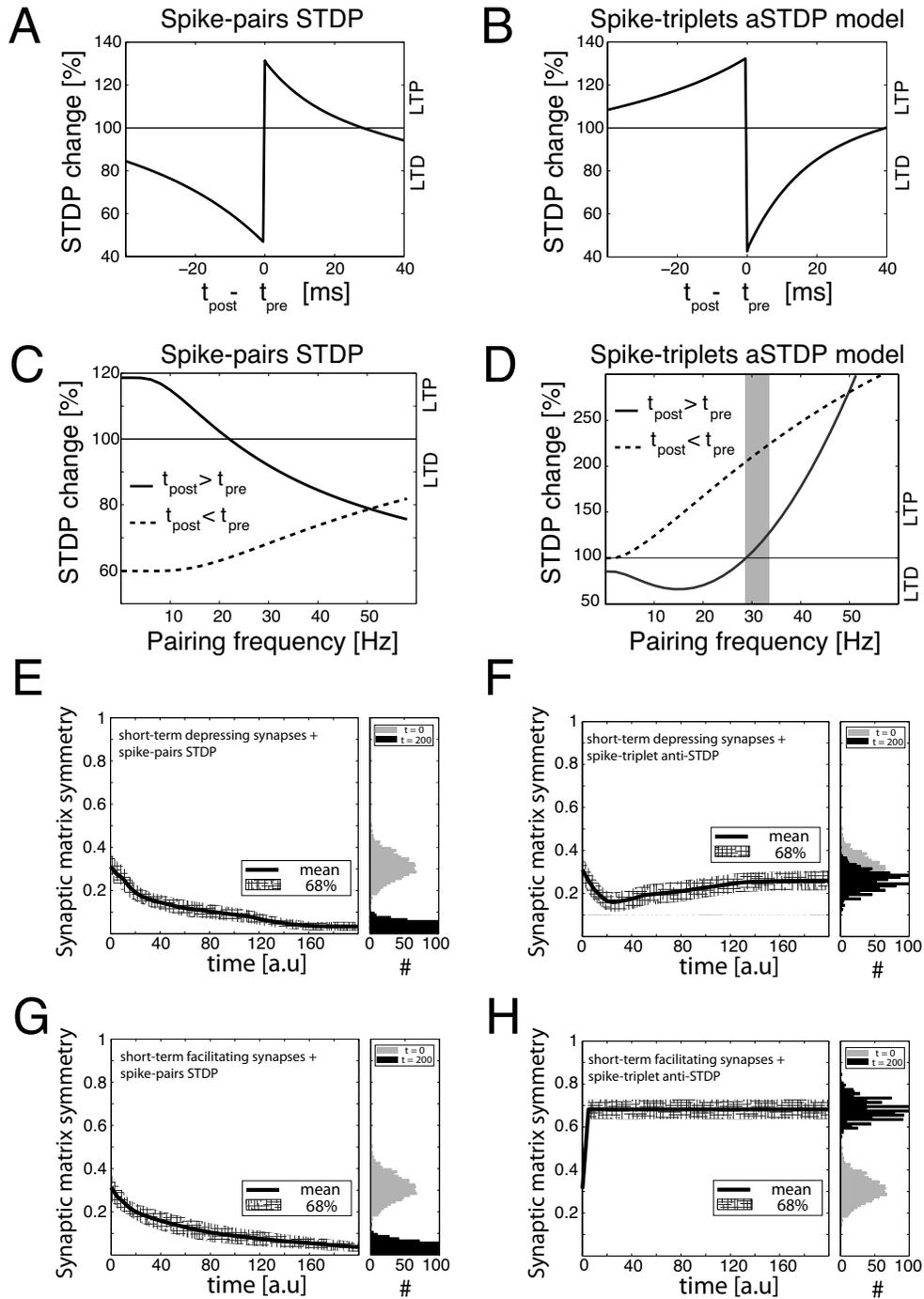}
\caption{{\bf STDP key-features for motif emergence}.  The {pair-based} STDP model, with temporal window shown in panel {\bf A} {and matching exactly \RefFig{fig:ToyStatistics}A}, exhibits a different frequency-dependency (panel {\bf C}) than the triplet-based STDP model (\RefFig{fig:ToyStatistics}B). Modifying the triplet-based STDP parameters to ad-hoc invert its temporal window (\eg as in anti-STDP, panel {\bf B}, compare to panel {\bf A}), yet leaves its frequency-dependency and the LTD-reversal (gray shading) unchanged (panel {\bf D}). Repeating the study of \RefFig{fig:ToyStatistics} with these two modified models, we find that i) {the pair-based} STDP fails to account for motifs emergence (panels {\bf E},{\bf G}, compare to \RefFig{fig:ToyStatistics}), while anti-STDP succeeds (panels {\bf F},{\bf H}, compare to \RefFig{fig:ToyStatistics}). Parameters: see the Supplemental Information. \label{fig:altSTDP}} 
\end{figure}

As a conclusive illustration of the simple mechanism for motifs emergence, we performed additional negative and positive control simulations (\RefFig{fig:altSTDP}), studying the impact of alternative formulations of STDP. By setting $A_3^+\ =\  A_3^- \ =\ 0$, and $A_2^+\ =\ 4.5\ 10^{-3}$ in \RefEq{STDPrule} and slightly modifying \RefEq{STDPindicators2} (see the Supplemental Information) one can reproduce the pair-based STDP model \cite{Song2000,PfisterGerstnerNIPS2006}. When parameters are adjusted so that this model has an identical spike-timing dependency (\ie compare \RefFig{fig:altSTDP}A to \RefFig{fig:ToyStatistics}A), the resulting frequency dependency {is completely different} (\ie compare \RefFig{fig:altSTDP}C to \RefFig{fig:ToyStatistics}B), showing no reversal of LTD into LTP at high frequencies. 
Vice versa, by inverting signs and swapping the values $A_2^+$, $A_3^+$, $A_2^-$, and $A_3^-$ in the same equations, it is possible to ad-hoc reverse the temporal dependency of STDP, as described experimentally for anti-STDP (aSTDP) \cite{BiRubin2005}, while leaving the frequency dependence roughly intact (\ie compare \RefFig{fig:altSTDP}B,D to \RefFig{fig:ToyStatistics}A,B) and featuring the reversal of LTD into LTP at high frequencies. Of course, this modified ``triplet rule" model should not be generalized as novel accurate description of aSTDP, since more data would be anyway needed to access its firing rate dependence.
Panels E-H of \RefFig{fig:altSTDP} repeat the simulations of \RefFig{fig:ToyStatistics}, and demonstrate that only when the frequency-dependence of STDP is realistic \cite{Sjostrom2001}, the {heterogeneity} in the network firing rates leads to the emergence of asymmetric connectivity motifs (compare \RefFig{fig:altSTDP}E,G or \RefFig{fig:altSTDP}F,H to \RefFig{fig:ToyStatistics}C,D).

\subsubsection*{Large homogeneous microcircuits}

Due to the simplicity of the mechanism disclosed above, we can analyse larger populations of neurons interacting via homogeneous and heterogeneous short-term plastic synaptic connections. Prior to running numerical simulations, we qualitatively investigate the generality of our results and the conditions for the asymmetry in the emerging firing rates. We carry out this analysis by a standard Wilson-Cowan firing rate description of neuronal networks dynamics (see \RefEq{mean_field_dyn}) \cite{DayanAbbott:2001}. This approach has been already extended to the case of short-term synaptic depression and facilitation (see \RefEq{mean_field_STP}) \cite{Tsodyks:2003,HolcmanTsodyks:2006,BarakTsodyks:2007}, long-term Hebbian plasticity \cite{DelGiudice2003}, or both \cite{DelGiudiceMattia2001}. This technique can be used as long as the characteristic times of long-lasting plasticity are much longer than those of neural and synaptic short-term dynamics, as in our case. 

\begin{figure}[htb!]
\centering
\includegraphics[scale=0.5]{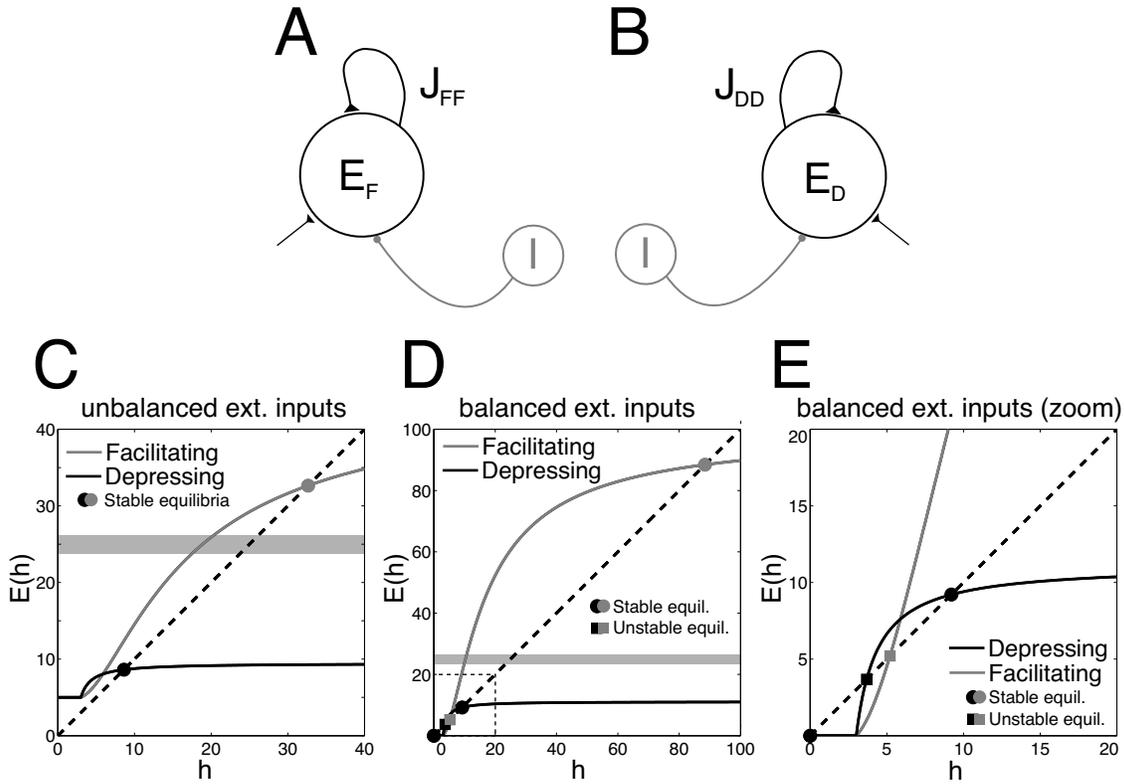}
\caption{{\bf Mean-field analysis of firing rate equilibria, in homogeneous networks without long-term plasticity}.  The firing rate of homogeneous recurrent networks, including short-term facilitating synapses ({\bf A}) or depressing excitatory synapses ({\bf B}), was studied by standard mean-field analysis. Average synaptic efficacies are indicated by $J_{FF}$ or by $J_{DD}$, correspondingly. Excitatory and inhibitory external inputs are modeled by a single term, taking positive, zero or negative values. A zero value corresponds to {\em balanced} excitatory/inhibitory inputs, while a non-zero value corresponds to {\em unbalanced} excitatory/inhibitory inputs.
The steady-state  firing rate (\ie $E$, in a.u.) are the roots of the equation $E(h)=h$ (see \RefEqs{mean_field_dyn},\ref{mean_field_STP} and the Supplemental Information), whose graphical solution is provided ({\bf C-E}), for facilitating (gray) or depressing (black) synapses, without long-term plasticity. Panel {\bf E} is a zoomed view of  {\bf D}. (Un)stable firing rate equilibria are indicated by filled circles (squares).  Networks with facilitating synapses fire at higher rates than networks with depressing synapses, as emphasized by the gray shading.
\label{fig:MeanField}}
\end{figure}

For the sake of simplicity, we initially omit STDP (\RefEq{mean_field_STP}), and consider recurrent networks of neurons connected by homogeneous synapses, facilitating (\RefFig{fig:MeanField}A) or depressing  (\RefFig{fig:MeanField}B). We sketch these networks as oriented graphs, where arrows represent connections with average efficacy $J_{FF}$ or $J_{DD}$, correspondingly. 
We do not explicitly include inhibition, which, nevertheless, does not qualitatively alter the validity of our results, as examined in the Supplemental Information. Instead, we distinguish whether the average external input to a generic neuron is zero, \ie referred to as {\em balanced} inputs ($I_{ext}\ =\ 0$ in \RefEq{mean_field_dyn}), or is set to a positive value, \ie referred to as {\em unbalanced} inputs, ($I_{ext}\ =\ 5$ in \RefEq{mean_field_dyn}). The population firing rates can be studied via standard methods as in dynamical system theory \cite{Strogatz:1994}, \ie analyzing the system of \RefEq{mean_field_dyn} and evaluating their steady-state solutions.

For instance, given a specific combination of external inputs and recurrent average synaptic efficacies, \ie 
$J_{FF}\ =\ J_{DD}\ =\ 4$ for unbalanced inputs and $J_{FF}\ =\ J_{DD}\ =\ 10$ for balanced inputs, the steady-state solutions of \RefEq{mean_field_dyn} provides the equilibrium firing rates. These can be then compared to the {\em critical} frequency of STDP (see \RefFig{fig:ToyStatistics}B), invoking the same mechanisms for motifs emergence proposed by Clopath {\em et al.,} (2010): if above the {\em critical} frequency, then LTP prevails and STDP will reinforce reciprocal synaptic efficacy; if below the {\em critical} frequency, then LTP or LTD will be determined by spike-timing information.
Panels C and D in \RefFig{fig:MeanField} illustrate graphically in the plane $E,\ h$, the actual location of the steady-state solutions of \RefEq{mean_field_dyn}. These are  derived by means of geometrical analysis techniques, similarly to the nullclines intersection methods \cite{Strogatz:1994}. In fact, at the equilibrium \RefEq{mean_field_dyn} can be rearranged so that its roots correspond to the intersections between the unitary slope line (dashed) and a given curve (continuous lines; see Supplemental Information). The shape and bending of such a curve depend on the external input $I_{ext}$ and on the SD parameters, so that the firing rates emerging in the recurrent networks of \RefFig{fig:MeanField}A-B can be compared to each other \cite{Tsodyks:2003,HolcmanTsodyks:2006,BarakTsodyks:2007}. The dynamical stability of each steady-state solution is displayed by a different marker symbols: circles for stable, squares for unstable equilibrium points. Similarly to \RefFig{fig:ToyStatistics}B, the approximate location of the STDP {\em critical} frequency has been indicated by a gray shading.
Although we did not chose the parameters of the rate models to quantitatively match the Integrate-and-Fire simulations \cite{AmitBrunel1997,DelGiudice2003}, we conclude that homogeneous facilitating networks generally fire at higher firing rates than depressing networks, for the same set of average synaptic coupling and external inputs, and when engaged in reverberating activity \cite{AmitBrunel1997}.

Mathematically the two networks share the same mean-field description, \RefEqs{mean_field_dyn}-\ref{mean_field_STP}, although their numerical parameters are considerably different (see Table \ref{tab:params}) \cite{Wang2006,Pignatelli2009}. Specifically, short-term depressing synapses have a much larger recovery time constant from depression ($\tau_{rec}$) than facilitating synapses. Facilitating synapses have instead a much larger recovery time constant from facilitation ($\tau_{facil}$) than depressing synapses.

Along these lines and by studying the asymptotic limits of the mean-field equations, we found heuristically that the dominating parameter is $\tau_{rec}^{-1}$. This sets an upper limit to the location of any possible firing rate of each network. Then, a network with low values of $\tau_{rec}^{-1}$, \ie where the time scale of recovery from depression is very long, would fire slower than a network with comparatively higher values of $\tau_{rec}^{-1}$, \ie where the time scale of recovery from depression is very fast or negligible. These two cases correspond to the depressing and facilitating networks, respectively (\RefFigs{fig:MeanField}A-B). 

These general considerations have been validated and confirmed in networks of $1000$  neurons, see Methods. In these simulations, neurons are connected with maximal synaptic efficacy $A=6\ pA$. In order to increase the biological realism, in these large scale simulations we introduce fluctuating random inputs to each neuron, mimicking background network activity  \cite{Tuckwell:1989}. Therefore, each neuron receives an uncorrelated noisy current, as well as a periodic wave-like stimuli (see Methods).
\RefFigs{fig:PopulationsMicroscopic}A-B display the count of the occurrence of  unidirectional {\em versus} reciprocal connectivity motifs. %
Similar to the small toy model, analysed in \RefFig{fig:ToyStatistics},  unidirectional depressing connections significantly outnumber the reciprocal depressing connections, while facilitating reciprocal connections prevail on  unidirectional facilitating connections.
As indicated by \RefFigs{fig:PopulationsMicroscopic}D-E, the distributions of the firing rates of the two networks feature the same  {heterogeneous distributions of firing rates}: networks of homogeneous depressing short-term plastic synapses fire at generally low rates, while networks of homogeneous facilitating synapses fire at higher rates. Finally, the symmetry index $s$, computed after a very long simulation run,  results in a value of $0.28$ for the depressing network and of $0.99$ for the facilitatory network.

\subsubsection*{Heterogeneous microcircuits}

We further study the more general case of a heterogeneous network (\RefFig{fig:PopulationsMacroscopic}A) with two subpopulations: one with  facilitating synapses, and the other with  depressing synapses. Arrows represent synaptic connections, with average efficacies $J_{FF}$, $J_{DD}$, $J_{FD}$, and $J_{DF}$. Synapses established within neuronal pairs that belong to the same subpopulation, share, by definition, the same SD properties, \ie short-term depressing or short-term facilitating, but not both simultaneously. On the contrary, synapses established within neuronal pairs that belong to distinct subpopulations have, by definition, heterogeneous SD properties. Thus a total of five categories of connectivity motifs are possible in this network: facilitatory reciprocal motifs, depressing reciprocal motifs, facilitatory unidirectional motifs, depressing unidirectional motifs, and reciprocal motifs with both facilitation and depression. For the first four categories, experiments support strong non-random occurrences \cite{Wang2006,Pignatelli2009}. For the case of reciprocal motifs with both facilitation and depression, no extensive experimental information has been published. Our results suggest that non-random occurrences of the first four categories arises from SD-STDP interactions, and predict that the last category should be largely underexpressed, compared to chance level.

We first examine the impact of STDP in a simplified two-neuron system, with one neuron projecting to the other via a single synapse. \RefFigs{fig:PopulationsMacroscopic}B,D,  show the long-term change in PSPs amplitude as a function of the pre- and postsynaptic firing rates, at that single synapse. Since in a heterogeneous network pre- and postsynaptic firing frequencies may differ, we swept the firing frequencies of the two neurons throughout all the possible combinations, within a realistic range. We study two cases: each presynaptic spike precedes the postsynaptic spike by $10\ msec$, \ie $t_{pre}<t_{post}$, or vice versa, \ie $t_{pre}>t_{post}$. We emphasise that only synapses established within neuronal pairs that belong to distinct subpopulations can experience heterogeneous pre- and postsynaptic firing rates. In this case however, the impact of spike timing information becomes negligible as soon as pre- and postsynaptic neurons fire at different frequencies. In the small minority of cases where this is not true, pre- and postsynaptic frequencies are integer (sub)multiples of each other, and a transient synchronization of spike times occurs periodically. In these circumstances, the timing information has a specific impact, as revealed graphically by bright or dark lines in the plots of \RefFigs{fig:PopulationsMacroscopic}B,D. In all other cases, overall plasticity profiles reflect the conventional associative Hebbian LTP/LTD and its consequences \cite{kempter1996,gerstner1996,DelGiudice2003}.

Intuitively,  the heterogeneous population of \RefFig{fig:PopulationsMacroscopic}A can lead to the emergence of connectivity motifs. To illustrate this point, we first ignore that  subpopulations might interfere with each other's firing rate. We  assume that the facilitating and depressing subnetworks would be still characterized by higher or lower firing rates, respectively, as previously presented for homogeneous networks. Then, the LTP/LTD maps of \RefFigs{fig:PopulationsMacroscopic}B,D suggest that such an initial asymmetry of emerging firing rates can be maintained indefinitely. The connections $J_{DF}$ are increasingly weakened and the connections $J_{FD}$ strengthen. The resulting configuration, sketched in \RefFig{fig:PopulationsMacroscopic}C, is stable. 
We tested and confirmed this statement under the mean-field hypothesis, by studying the dynamics of \RefEqs{mean_field_dyn}, \ref{mean_field_STP}, and \ref{mean_field_STDP}, in addition to computing their equilibria. \RefFig{fig:PopulationsMacroscopic}E-F display the mean firing rates of each subnetwork, and the initial time course of synaptic efficacies, with $J_{DF}$ progressively becoming weaker. This is true only when inter-population efficacies, \ie $J_{FD}$ and $J_{DF}$, are initialized to slightly weaker values than the intra-population efficacies, \ie $J_{FF}$ and $J_{DD}$. Such an initial difference between inter-population and intra-population couplings could however emerge from fully homogeneous couplings, \ie $J_{FF}\ =\ J_{DD}\ =\ J_{DF}\ =\ J_{FD}\ =\  1$, as demonstrated in the Supplemental Information.

\begin{figure}[htb!]
\centering
\includegraphics[scale=0.52]{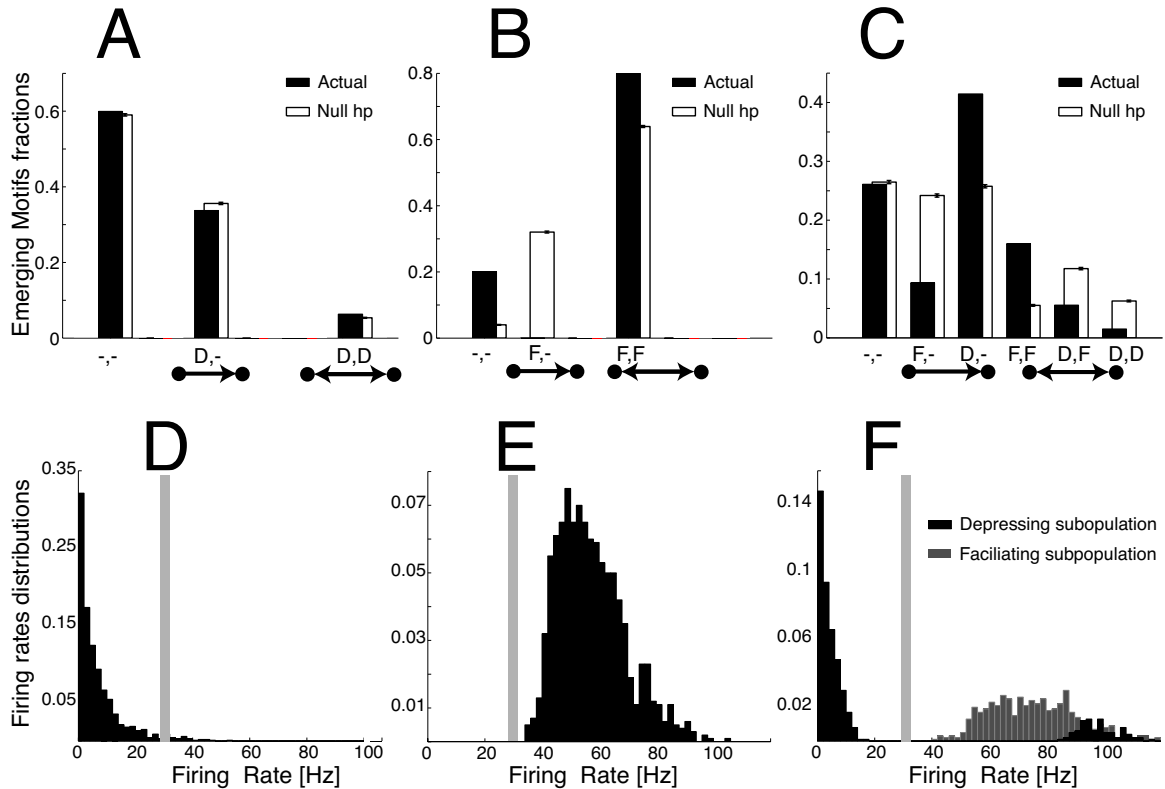}
\caption{{\bf Results from numerical simulations of large recurrent networks of model neurons with short- and long-term plasticity.}
Homogeneous and heterogeneous recurrent networks made of $1000$ Integrate-and-Fire model neurons were numerically simulated, under identical conditions.
Panel {\bf A} shows the comparison of the emergence of weak or no connectivity pairs (indicated as "-,-"), of unidirectional strong connectivity pairs  ("$\rightarrow$", "D,-"), and of reciprocal strong connectivity pairs ("$\leftrightarrow$", "D,D") for a homogeneous network of neurons connected by depressing synapses: strong unidirectional depressing connections significantly outnumber reciprocal depressing ones. The fractions of emerged motifs (black) is significantly different than the null-hypothesis (white) of random motifs occurrence.  Panel {\bf B} repeats this quantification for a homogeneous network with facilitating synapses: strong connections are only found on reciprocal connectivity pairs ("$\leftrightarrow$", "F,F") and all emerging motifs are non-random. Panel {\bf C} repeats the same quantification for a heterogeneous network with both short-term facilitating and depressing synapses. Emerging motifs display highly non-random features and confirm that reciprocal facilitatory motifs ("$\leftrightarrow$", "F,F") outnumber unidirectional facilitatory motifs ("$\rightarrow$", "F,-"), and that unidirectional depressing motifs ("$\rightarrow$", "D,-") outnumber reciprocal depressing motifs ("$\leftrightarrow$", "D,D"). Panels {\bf D-F} display the steady-state firing rate distributions, corresponding to homogeneous depressing, homogeneous facilitating, and heterogeneous networks respectively. The plots confirm that  {heterogeneity} in connectivity motifs is accompanied by bimodal firing rates above and below the {\em critical} frequency, represented here by a grey shading.
\label{fig:PopulationsMicroscopic}} 
\end{figure}

\begin{figure}[htb!]
\centering
\includegraphics[scale=0.65]{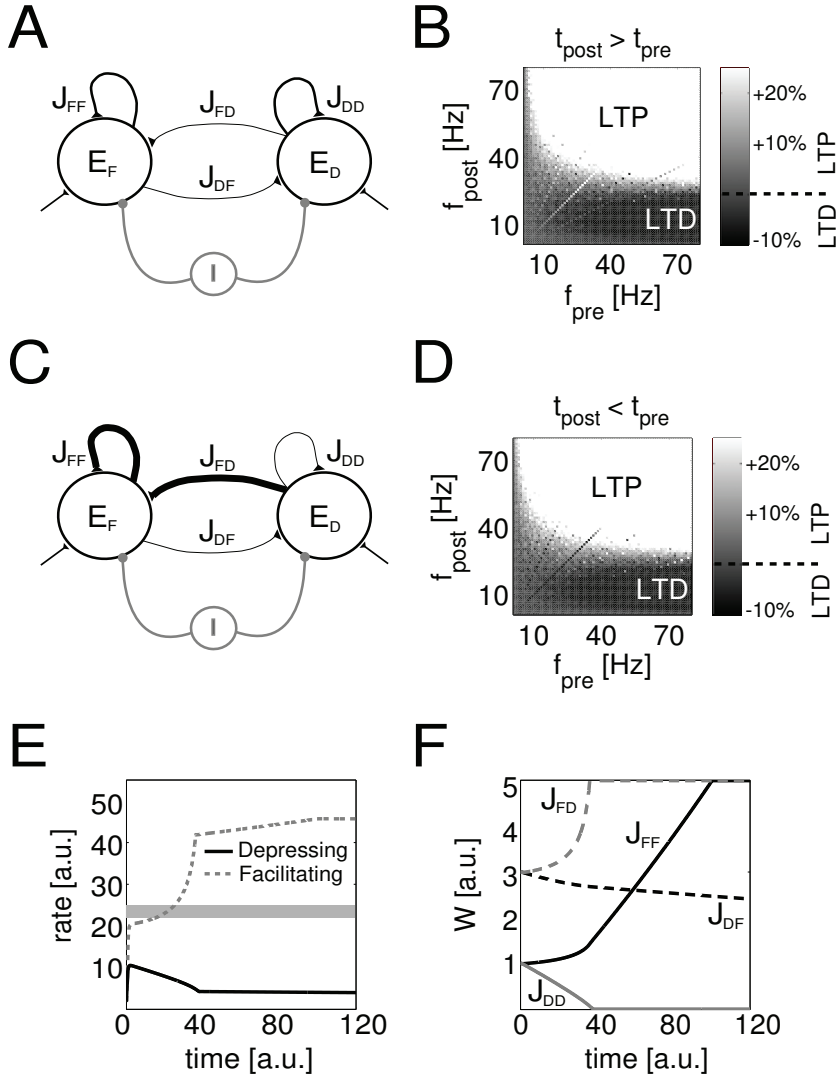}
\caption{{\bf Mean-field simulation of a heterogeneous network with short- and long-term plasticity}. The firing rate evolution of a heterogenous recurrent network, including both short-term facilitating and depressing synapses ({\bf A}), was estimated by numerically solving the corresponding mean-field equations. The average synaptic efficacies among and across populations, indicated by $J_{FF}$, $J_{DD}$, $J_{FD}$, and $J_{DF}$,  undergo long-term modification. Panels {\bf B,D} show the  long-term changes of an isolated synapse (decoupled from recurrent interactions) depending on pre- and postsynaptic spike timing (\ie $t_{pre}$, $t_{post}$) and frequencies (\ie $f_{pre}$, $f_{post}$). When $f_{pre}$ and $f_{post}$ are varied independently, long-term potentiation (LTP) and depression (LTD) emerges as in associative Hebbian plasticity. This suggests that $J_{FF}$ and $J_{FD}$  will become significantly stronger than $J_{FD}$ and $J_{DD}$ ({\bf C}) and that such a configuration will be retained indefinitely. This was confirmed by  simulations ({\bf E-F}) plotting the  temporal evolution of the firing rates $E_F$ (black trace) and $E_D$ (gray trace), and of the synaptic efficacies ({\bf F}). The  {heterogeneity}  occurs by separation of emerging firing rates (\RefFig{fig:MeanField}), as emphasized by the grey shading.  Parameters: $I_{ext}=5$, $\tau=10\ msec$, with initial conditions  $J_{FF}=J_{DD}=3$, and $J_{DF}=J_{FD}=1$ (see Supplemental Information for alternatives).
\label{fig:PopulationsMacroscopic}}
\end{figure}

We further confirmed that the synaptic configuration, indicated by the mean field models, is present in large microscopic numerical simulations, as in \RefFig{fig:PopulationsMicroscopic}C (see Methods). These simulations involve $1000$ identical Integrate-and-Fire units, subdivided in two subpopulations of equal size, with their anatomical connectivity set to $80\%$ of all possible connections, similar to the homogenous case. The maximum synaptic efficacy is set to $A=12\ pA$ and the initial synaptic connections $W_{ij}$ are randomly initialized. As in the mean-field model, the inter-population terms $W_{ij}$ are initialized to weaker values than the intra-population terms (but see Supplemental Information for alternatives). Each neuron receives an uncorrelated background noisy current as well as periodic wave-like stimuli, similar to the homogeneous case.
As indicated by \RefFig{fig:PopulationsMicroscopic}F, the distributions of the spiking frequencies of individual neurons of the two subnetworks feature the same  {heterogeneous distribution of firing rates} as in the previously studied homogeneous networks: subnetworks of depressing short-term plastic synapses fire at generally low rates, while subnetworks of  facilitating synapses fire at higher rates. The location of the {\em critical} firing frequency for the STDP is represented again as a grey shaded area. 

Results in \RefFig{fig:PopulationsMicroscopic}C show all the possible synaptic combinations. As in the data of Wang {\em et al.} (2006)\nocite{Wang2006}, reciprocal motifs are significantly co-expressed with facilitatory synapses and  unidirectional motifs with depressing synapses. The actual motif count is compared to the null-hypothesis of having statistical independence between the connection occurrence within a pair of neurons, estimated at a $95\%$ confidence interval upon the same hypothesis of {\em Bernoulli} repeated, independent, elementary events. The frequency $Q$ of observing a connection between two neurons, regardless of its SD properties, is first estimated by direct inspection of the connectivity matrix $[W_{ij}]$. Then the conditional occurrence frequencies of a facilitatory synapse $Q_F$ and of a depressing synapse $Q_D\ =\ (1-Q_F)$ are computed, given that a connection exist between two neurons. The null hypothesis for each possible combination is given by standard probability calculus, under the hypothesis of independence of the identical events. For instance, the occurrence frequency of observing by chance no connections within a neuronal pair is $(1-Q)^2$, while the occurrence of observing by chance a reciprocal motifs with mixed depressing and facilitating properties is $2 \ (Q^2\ Q_F\ Q_D)$. Finally, in this example, the symmetry index $s$, computed after a very long simulation run, resulted in a value of $0.18$ for the depressing subnetwork and of $0.66$ for the facilitation subnetwork.

\subsubsection*{Microcircuits with overlapping SD properties}

In the heterogeneous network of \RefFig{fig:PopulationsMacroscopic}, as well as in the homogeneous networks, we make the assumption that the SD profile is determined primarily by the identity of the projecting neuron. This has been experimentally found in the olfactory, visual, and somatosensory cortices as well as in other brain areas \cite{BowerHaberly1986,Stratford1996,ReyesSakmann1999,Geracitano2007}.
Nonetheless, the assumption on the projection-cell specificity can be removed in order to theoretically explore the impact of SD heterogeneity across distinct synaptic connections established by the same presynaptic neuron \cite{MarkramWangTsodyks1998}.

We assumed that a generic neuron had a certain probability $p_D$ of establishing a short-term depressing synapse with a target neuron, and a probability $1-p_D$ of establishing a short-term facilitating synapse with another one. In this case, individual neurons are still indistinguishable and their mean-field synaptic input can be described mathematically \RefFig{fig:MixedMeanField}A (see Supplemental Information).

For small values of $p_D$, the emerging firing rates approximate those of a network of facilitating synapses, while for large values of $p_D$ the firing rates behave as for a network of depressing synapses. In other words, the mixed networks behave dynamically as an intermediate case between two extremes. This result is quantified in figure \RefFig{fig:MixedMeanField}B, where the location of the stable equilibrium points has been analysed under the mean-field hypotheses and plotted as a function of $p_D$, for different external inputs regimes. The qualitative location of the {\em critical} firing frequency for the STDP is represented as a grey shaded area. In this situation, and as an explicit consequence of the lack of any structure, \ie compare \RefFig{fig:MixedMeanField}A with \RefFig{fig:PopulationsMacroscopic}A, STDP fails to discriminate individual connections within the network, but rather shapes them as reciprocal or as  unidirectional motifs, depending on the particular choice of $p_D$.%
\newpage
\section*{Discussion}

{Our results indicate that time- and frequency-dependent STDP mechanisms may be responsible, through internally generated spiking activity in recurrent network architectures, for the observation that excitatory neurons connected by short-term facilitating synapses are more likely to form reciprocal connections, while neurons connected by  short-term depressing synapses are more likely to form unidirectional connections. More specifically:}
\begin{enumerate}
\item
{the internally generated firing rates in model networks with facilitating connections are higher than in networks with depressing connections, under identical background inputs;}

\item 
{neurons, participating to such an internally generated activity, are likely to form bidirectional connections with each others when firing at sufficiently high rates, reflecting the ``Hebbian" mode of STDP; neurons firing at low rates are likely to form unidirectional connections, reflecting the temporally asymmetric ``pre-post" temporal mode of STDP;}

\item 
{once formed, these connectivity motifs persist and self-sustain themselves through the internal firing regimes of the network, from which the motifs emerged from;}

\item 
{externally generated inputs, strongly depolarizing or strongly hyperpolarizing individual neurons, when prevailing over internally generated activity, may lead to opposite motifs emergence consistent with Clopath {\em et al.} (2010)\nocite{Clopath2010}.}

\end{enumerate} 

\subsection*{Relationships to previous work and additional mechanisms}

The impact of long-term associative synaptic plasticity in recurrent networks of spiking neurons has been earlier studied by many investigators,  who proposed mechanisms  for the unsupervised formation of stimulus-driven dynamical attractors of network activity, in the context of  working-memory states \cite{DelGiudice2003}. Within the same aims, the interactions between long-term plasticity and SD were also already partly explored, both in numerical simulations and in mean-field descriptions \cite{DelGiudiceMattia2001}. Here, we focused on specific long-term plasticity mechanisms (STDP) \cite{MarkramGerstnerSjostrom2011} to study the emergence of network structure \cite{Morrison2007,Gilson2010,Gilson2009,Kunkel2011} and connectivity motifs \cite{Clopath2010,BourjailyMiller2011}. To the best of our knowledge, this is the first time that SD is viewed as key element for the emergence of network connectivity, and shown to capture,  to a certain extent, the experimental observations.

Our study builds upon assumptions of Clopath {\em et al.} (2010)\nocite{Clopath2010}: {\em in silico} activity-dependent (dis)appearance of a connection occurs in terms of a competition among  ``strong" links in a ``sea" of weak synapses, which are undetectable by cellular electrophysiology. There is still no general agreement on whether STDP and its variants can entirely account for developmental wiring, fine-tuning, or remapping of microcircuits connections \cite{SongAbbott2001}. However, the role of STDP for connectivity development has been positively discussed for systems where information is present at fast time scales \cite{ButtsKanold2010}, as it might be relevant for the neural code of cortex and olfactory microcircuitry. In addition, the STDP ``triplet" rule  proposed by Pfister and Gerstner \cite{PfisterGerstner:2006}, which is the central ingredient for this work, captures the behaviour of  developmental plasticity models \cite{Gjorgjieva2011} widely used in classic studies of connectivity development \cite{Cooper2004}. As a consequence, our results are built upon well-known phenomenological models, which have been shown to have excellent agreement with experimental data sets \cite{TsodyksMarkram:1997,PfisterGerstner:2006,Clopath2010}. We have, therefore, neglected more accurate biophysical descriptions of SD \cite{Fuhrmann2002,Fuhrmann2004,Loebel2009,PanZucker2009} and STDP \cite{Rubin2005,GrapunerBrunel2012}, aiming at a minimalist  description of neuronal excitability, synaptic transmission, and synaptic plasticity. The advantage of our approach is that the functional consequences of the interactions between SD and STDP could be analysed by standard mean-field analysis \cite{Renart2004,BarakTsodyks:2007}.

Our study concludes that the symmetry of the connections may be influenced by the internally generated spatio-temporal correlations of neuronal firing \cite{Vogels2005}. Therefore, connectivity motifs might not emerge exclusively from correlations governed  by the external inputs \cite{Clopath2010}. Both internally generated and externally imposed correlations are likely to play a role 
and therefore the results of  Clopath {\em et al.} (2010) still hold. For instance, in a microcircuit connected by purely depressing synapses, external activity can still induce heterogeneity in the firing rates. If a subset of neurons {was strongly depolarized by external selective inputs, then units  would} fire above the critical frequency, {and} non-random bidirectional connectivity might equally emerge {(see \RefFig{fig:asinClopath}). Similarly, strong hyper polarization by external selective inputs, would cause neurons to fire below the critical frequency, even though neurons are connected by recurrent facilitating synapses (not shown).}. 

Out of many possible features affecting internally generated activity, such as cellular excitability, architecture of long-range connections, and SD properties, here we focused on the last one. 
We note however that the dataset of Wang {\em et al.} (2006)  contains additional information on cell excitability and its correlation to motif emergence. Cells displaying accommodating discharge patterns (\ie exhibit spike frequency adaptation) are found to establish mostly unidirectional motifs. Instead, cells displaying nonaccommodating discharge patterns establish mostly reciprocal motifs \cite{Wang2006}. 
A differential expression of spike-frequency adaptation currents results in (non)accommodating discharge patterns. This is known to affect the firing rate distributions in recurrent model networks \cite{Gigante2007,Rauch2003,LiuWang2001}, and generally lower the location of steady-states stable equilibria (\RefFigs{fig:MeanField}C-E).
For the model parameters employed here, we did not observe differences when repeating all our simulations without  spike-frequency adaptation. Adaptation currents in the model are not strong enough to reduce substantially the steady-state firing rates of \eg the facilitatory subnetworks below the {\em critical} firing rate for STDP, and violate our conclusions.
On the contrary, as spike-frequency adaption is reported by Wang {\em et al.} in neurons participating to unidirectional motifs, we speculate that it participate jointly with depressing SD in maintaining STDP in its correlational ``pre-post" mode: once more this determines the emergence of unidirectional connectivity motifs. Non-accommodating discharge patterns  would by definition not interfere with the output firing rates, in networks connected by facilitating synapses and bound to emerge reciprocal motifs, by the ``Hebbian" mode of STDP.

Our study addresses the emergence of motifs in the cortical pyramidal microcircuitry, but as preliminary data collected in the olfactory bulb become available, our framework could tentatively model a common principle for synaptic wiring. Long-term and short-term plasticity has been experimentally found among olfactory mitral cells, \cite{PimentelMargrie2008,Pignatelli2009}, and STDP was reported in the rodent and insect olfactory systems \cite{Cassenaer2012,Gao2009}, and invoked at the output of mitral cells to explain decorrelation of sensory information \cite{Linster2010}. We, therefore, speculate that similar mechanisms for the emergence of connectivity motifs are common in both the cortical and olfactory microcircuitry.

Our investigation of heterogeneous network architectures revealed that STDP and SD led to a stable configuration of developing connections, compatible to the experimental observations. Under the simplified hypotheses that we adopted, the configurations of connectivity motifs and their relationship to the underlying neuronal collective dynamics are stable. Our results have been grounded on both numerical simulation results and mean-field analysis. The latter, however, ignored spike-timing correlations between spike-trains. It could thereby only assess the conditions for the emergence of strong connections. While this is sufficient to anticipate the emergence of reciprocal connectivity motifs, its converse does not per se implies the emergence of unidirectional motifs. In such a case, connections would for instance tend to weaken or to disappear. It is only through the numerical simulations of the full network of spiking neurons that we could show that the causal spike-timing information, if present, gives rise to unidirectional connectivity via STDP mechanisms. 

We also note that the major difference in SD properties, which accounts for motifs emergence, is the heterogeneity of the time constant representing the short-term depression recovery $\tau_{rec}$. In this respect, our results and conclusions would be qualitatively unchanged by replacing facilitating synaptic properties with linear, \ie non-depressing, properties. Along these lines, we hope that our results might prompt new experiments on connectivity motifs. We predict that the value of $\tau_{rec}$, in a pair of connected neurons, should be inversely correlated to the occurrence frequency of reciprocal motifs.

\subsection*{Simplifying assumptions}

Among the key simplifying hypotheses of this paper, we note that STDP was assumed to scale only the SD parameter $G$, while leaving the parameter $U$ unaltered \cite{Buonomano1999}. This choice is consistent to what has been reported at distinct synapses of the central nervous system, although it might not be representative of all cortical areas  \cite{MarkramTsodyks1996}. Although the debate on pre- and postsynaptic expression of both SD and STDP is fierce, our choice of SD and STDP interaction is in part an arbitrary hypothesis, and it serves as a first solid ground for our conclusions. Enabling STDP to modify the parameter $U$ would have partly altered in an activity-dependent manner, the SD profile of a synapse. This would have made isolating SD contribution more complex, and relating our findings to previous theoretical works \cite{Morrison2007,Gilson2010,Gilson2009,Clopath2010,Kunkel2011} less straightforward. 

The model itself is purely phenomenological, and does not capture biophysical details, but rather the interaction of SD and STDP via a set of variables locally known to the synapse. It does however maintains the desirable compatibility with experimental data. In addition, explicit and systematic details on STDP at excitatory facilitatory synapses in the ferret prefrontal cortex are currently scarce  \cite{Wang2006}. While more efforts, both experimental and theoretical, should be undoubtedly devoted in these directions, the hypothesis of scaling $G$ and not $U$ remains a simplifying assumption, in the view of lacking of a systematic understanding on how STDP affects all the parameters of the SD model \cite{MarkramPikus1998}.

The presence of two classes of excitatory cortical synapses \cite{Wang2006}, with a distinct set of SD parameters such as $\tau_{rec}$, $\tau_{facil}$, and $U$, prompted us to consider in our model the existence of two classes of excitatory neurons. The consequence of having these classes within the same microcircuitry, as well as their postsynaptic target preferences for connection, are crucial issues that deserve more experimental and theoretical efforts, but see however \cite{BarakTsodyks:2007,Mongillo2008}. We thus explicitly neglected heterogeneity in the target synaptic preference of neurons, when expressing a facilitating or depressing SD profile \cite{MarkramWangTsodyks1998}. We instead considered homogeneous and heterogeneous neuronal populations, where the facilitating or depressing SD profile was determined by the projecting neuron. 
Although such a scenario for SD seems realistic for synapses between principal neurons, it might not be accurate for all the synapses in neocortical microcircuits.

The mixed microcircuitry of \RefFig{fig:MixedMeanField} and its preliminary analysis are an example of our attempt to relax these assumptions. Depending on the probability of establishing a  facilitating (or depressing) connection, fully mixed networks were shown to behave as a continuum between the two extremes of the homogeneous populations studied earlier, \ie networks of all facilitating or all depressing connections. These mean-field predictions were further confirmed in numerical simulations of large networks (not shown).  In these cases, SD-STDP interactions will not necessarily lead to the desirable  {heterogeneity} for the connectivity motifs, as the  {heterogeneity in the emerging firing rates} does not occur. Structured local microcircuit architectures, with heterogeneous single-cell type, number of afferents, and firing rates are the obvious topics for a future study. We believe that the simple case of the heterogeneous two-subpopulation network studied in this work was a first necessary step towards the highlighted future directions.

Another point that deserves some discussion is our choice for a structured, foreground periodic stimulation, employed in all simulations with more or less intense asynchronous background activity. In the toy microcircuit (\RefFig{fig:ToyNetwork}), the cyclic stimulus was needed to demonstrate the emergence of a large number of asymmetric connections. It is, however, neither a strict requirement for our interpretation of reciprocal and unidirectional motifs formation, nor a necessary condition for their unbalance. This cyclic, deterministic, stimulus is an oversimplified way to invoke the temporal coding strategy of \cite{Clopath2010}. It was chosen to unambiguously relate unidirectional pairwise motifs formation to spiking activity. Without such an input, and under the presence of strong uncorrelated background activity, unidirectional motifs would tend to be very sparse.  We underline that our intention was to expose to the very same conditions networks identical in every aspects but in the synaptic SD, and validate the   {heterogeneity} conditions of the connectivity matrix.

\subsection*{Limits of our approach}

Our proposed mechanism for non-random pattern emergence is based on the sole interactions between STDP and SD. Obviously, it is unlikely that these mechanisms operate independently of other synaptic phenomena. Homeostatic plasticity could for instance continuously rescale synaptic efficacy and make SD heterogeneities less predominant in determining connectivity motifs. In the lack of {\em a priori} experimental information, we chose the maximal synaptic efficacy $A$ to share the exact same value for all the microcircuits we examined, in order to ensure a fair comparison.
With all its limitations, our proposal may still provide a simple working hypothesis on one component underlying the emergence of connectivity, linked to short-term synaptic dynamics, along the same lines of the theory proposed by Clopath {\em et al.} (2010)\nocite{Clopath2010}. Its validity could be challenged by experiments that interfere and probe the emergent firing activity, \eg in local {\em in vitro} cultured microcircuits with known synaptic properties \cite{BiPoo1998}. Our framework might be also useful for investigating further structure-function relationships at the subcellular level, by altering the synaptic machinery, or by employing (future) genetically-encoded fluorescent reporters of synaptic efficacy and dynamics. The use of optogenetics and genetically encoded neuronal voltage- and calcium-sensors, may lead to experimental validation or falsification of our hypothesis, which might directly contribute to understand short- and long-term plasticity interactions.

Further, the results of our simulations do not automatically capture the over-expression of reciprocal connectivity cortical motifs over the unidirectional ones  \cite{Song2005}. This is violated in the case of short-term depressing synapses, where the expression of reciprocal motifs is clearly at chance level \RefFig{fig:PopulationsMicroscopic}A, but it is however not the case for \RefFig{fig:PopulationsMicroscopic}B. Clopath {\em et al.} (2010) model these aspects by employing different neuronal coding strategies, \ie time- or rate-codes.
In our work, in order to isolate the  {heterogeneity} component of SD properties, we considered a common type of spatio-temporal activation and ignored, for the sake of simplicity, heterogeneous neural coding strategies. We are however confident that including more complex network architectures, as well as more specific spatio-temporal correlations, as in Clopath {\em et al.} (2010), additional non-random features could be explained.

{In our results, we use oversimplified network architectures to illustrate that two recurrent excitatory networks, identical in any aspect apart of their synaptic properties, evolve different connectivity motifs. We simulated and studied these models analytically, and we identified the mechanism behind the heterogeneity as frequency-related. It would not be an obvious easy task to explain simultaneously all the properties of cortical connectivity, as experimentally observed \cite{Wang2006}. The choice of our homogeneous architecture, the lack of cell diversity, and the isolated evolution of neuronal dynamics,  considerably differ from the actual cortex as well as from the variety of internal and external stimuli it receives during development and during lifetime. Nevertheless, the choice of our protocol serves in demonstrating a potential mechanisms behind the motif formation, for facilitating and depressing synapses.}

We would like to emphasise that our theory refers only to one of many possible, perhaps competing, mechanisms that contribute to stereotypical motifs emergence. Alternative explanations and a causal demonstration of the key ideas we suggest, remain to be provided. It might be of interest exploring to which extent developmental changes in SD, such as the switch from depression into facilitation at synapses  between layer 5 pyramidal neocortical neurons \cite{ReyesSakmann1999}, occurring after postnatal day (P) 22, are mirrored by changes in motifs statistics. 
For marginal pair-wise probability of connection, Song {\em et al.} (2005)\cite{Song2005} report no significant dependence on age, but  provide no systematic characterization of motifs statistics beyond P20. 

It may be also possible to attempt a chronic manipulation of the firing rates of neuron (sub)populations, by pharmacologically altering synaptic profiles, e.g., modulating postsynaptic receptor desensitisation, changing the presynaptic probability release, or interfering with neurotransmitter recycling.
As future directions, more complex heterogeneous anatomical architectures and single-cell properties should be incorporated within the same computational modeling framework. Very specific, non-random initial architectures, \eg small-world and scale-free \cite{Barabasi2005}, could be explored, extending our results towards other aspects that determine reciprocal or unidirectional motifs, possibly beyond the firing levels and towards, \eg, the density of hub nodes, ranking orders, heavy tails in neighbours distribution, etc.

How critical is the STDP ``triplet" rule for the validity of our results? Any STDP model that is able to capture the high frequency effect on plasticity, as revealed by the experiments of Sj\"ostr\"om {\em et al.} (2001)\nocite{Sjostrom2001},  will reproduce our results if combined with short-term facilitating and depressing synapses. {Not all STDP models are consistent with the data of Sj\"ostr\"om }. For instance, Froemke and Dan \cite{FroemkeDan2002} proposed an STDP model capturing pre- and postsynaptic frequency effects. This model predicts that  in high frequencies no reversal of LTD into LTP takes place, an inversion that is critical for the mechanism we propose for motifs emergence. Interestingly, however, a generalization of the ``triplet" rule that we adopt here is able to capture the experimental data of Froemke and Dan (2002)\nocite{FroemkeDan2002} \cite{ClopathGerstner2010}.
{We also note that pair-based STDP models may still support the emergence of reciprocal connectivity motifs and replicate our results (not shown). In fact, by adjusting the numerical values of its parameters (e.g.,  $A_2^+ = 2 A_2^-$ in  Eqs. S39 and S42, see Supplemental Information), LTP can be made prevailing over LTD above a frequency similar to the critical value of the triplet-model \cite{SongAbbott2001,Kepecs2002}. Nevertheless, when the pair-based STDP parameters are chosen to match the 10Hz temporal window of the triplet-STDP model employed here (compare \RefFigs{fig:ToyStatistics}A and \ref{fig:altSTDP}A), the frequency-dependent profile observed by Sj\"ostr\"om {\em et al.} (2001)\nocite{Sjostrom2001} is not reproduced (compare \RefFigs{fig:ToyStatistics}B and \ref{fig:altSTDP}C) and SD-STDP interactions do not lead to the same results (compare \RefFigs{fig:ToyStatistics}C,D and \ref{fig:altSTDP}E,G). This served here as a negative controlsk for our results.}
We also remark that excluding heterogeneity in spike propagation delays made our analysis simpler. It is known that the distribution of delays considerably affect emerging network states, besides having an obvious effect on STDP \cite{MattiaDelGiudice2004}. In this respect, not only STDP and SD interactions might be altered by propagation delays, but the collective dynamical properties of recurrent networks could also vary. Once more, for the sake of simplicity we chose to consider a minimalist scenario, setting the ground for future, more extended, investigations.

Finally,  we underline the great value of the availability of physiological information accompanying anatomical connectivity. These complementary data-set contains precious statistical information regarding the expression of microcircuit motifs, which we are starting to recognize \cite{Song2005,Perin2011}. We believe that computational modeling is in this context a very powerful tool to explore additional hypotheses and challenge further theories.

\appendix

\section*{Acknowledgments}

We are grateful to C. Clopath, J.-P. Pfister, M. Pignatelli, A. Carleton, P. Dayan, A. Loebel for discussions and comments.

 \section{Supplementary Text}

\renewcommand{\theequation}{S\arabic{equation}}
For the sake of illustration of the standard mathematical techniques employed in the 
main text, we consider two extreme simplifications of the mean-field description of a recurrent networks with short-term plastic synapses 
(Eqs. 9-10; see also \cite{BarakTsodyks:2007}, and references therein): the  predominance of depressing mechanisms or the predominance
of facilitatory mechanisms. We analyze these two cases and specifically  study the stability of  dynamical equilibria. These simplified models 
and their analysis have been already provided in the literature, with varying degree of details.  We further comment on the analysis of the full 
model of short-term synaptic plasticity and comment on the impact of adding recurrent inhibition to our scenario. We then consider and
numerically analyse the mean-field description of a mixed population, where a generic neuron may establish simultaneously depressing and
facilitating synapses to its targets. 
We finally examine and partly relax the necessary conditions for a heterogeneous network of two interacting subpopulations to display emergence of connectivity motifs. 
For the case of a random matrix (\ie as a null hypothesis), we provide
some of the statistical properties of the symmetry measure we adopted to quantify the connectivity motifs to derive a confidence measure. In the last section, we provide model details for the {pair-based}, STDP as well as for the anti-STDP ``triplet" model, discussed in the main text.

\subsection{A single population with depressing synapses}
The Eqs. 9-10 of the main text can be simplified under the hypothesis of very fast recovery from facilitation. This is the same of
assuming that $\tau_{facil}$ is very small and that $u\ =\ U$ does not vary in time, as a consequence. In this case, only short-term depression is
modifying the synaptic efficacy, so that the mean-field firing rate dynamics of a large recurrent network of excitatory neurons, can be described by
a system of two dynamical equations \cite{Tsodyks:2003}:

\begin{equation}
  \left\{
    \begin{array}{l}
     \tau\ \dot{h}\ =\ -h\ +\ A\ U\ x\ E\ +\ I_{ext}\\ [3mm]
     \dot{x}\ =\ \left(1\ -\ x\right) / \tau_{rec}\ - U\ x\ E\\[3mm]
    \end{array}
  \right.
\label{depressing_mean_field}
\end{equation}

These are coupled non-linear differential equations that can be analyzed by standard methods of dynamical systems theory \cite{Strogatz:1994}.
We will consider here the derivation of the equilibria for $h$ and $x$, and indicate how their stability was assessed. By definition of equilibrium, 
$\dot{h}\ =\ 0$ and $\dot{x}\ =\ 0$ for those points (also called {\em fixed} points). Substituting these conditions in \RefEqs{depressing_mean_field}, we
obtain two non-linear algebraic equations,

\begin{equation}
  \left\{
    \begin{array}{l}
     h =\ A\ U\ x\ E\ +\ I_{ext}\\ [3mm]
     x\ =\ 1\ / \left( 1\ +\ U\ \tau_{rec}\ E\right)\\[3mm]
    \end{array}
  \right.
\label{depressing_mean_field_steadystate}
\end{equation}

Using the second equation to replace $x$ as it appears in the first, we get an implicit equation in the unknown $h$:

\begin{equation}
    h =\ A\ U\ E\ / \left( 1\ +\ U\ \tau_{rec}\ E\right)\ +\ I_{ext}
\label{phase_space}
\end{equation}

\RefEq{phase_space} can be solved numerically (\eg by the Newton-Raphson method, \cite{Press:2007}), given a specific set
of parameters $A$, $U$, $\tau_{rec}$, and $I_{ext}$.  Alternatively, its solution(s) can be interpreted
graphically as the intersection(s) between two functions of $h$, $F_1(h)$ and $F_2(h)$, in the cartesian plane,

\begin{equation}
  \left\{
    \begin{array}{l}
     F_1(h)\ =\ h\\ [3mm]
     F_2(h)\  =\ A\ U\ E\ / \left( 1\ +\ U\ \tau_{rec}\ E\right)\ +\ I_{ext}\\[3mm]
    \end{array}
  \right.
\label{graphical_solution}
\end{equation}

\noindent with $F_1(h)$ is the unitary slope line (see \RefFig{fig:F2plot}). Retaining the graphical interpretation, it is possible to appreciate intuitively the existence of the equilibrium points and their dependence on the parameters, as explored in \RefFig{fig:F2plot}. To this aim it is useful, prior to plotting $F_2(h)$, to analytically determine some of its mathematical properties, such as the asymptotic limits and derivatives.  

We observe that, by definition of $E\ =\ \left[\alpha\ (h - \theta)\right]_+$, $F_2(h)$ is zero for values of $h$ lower than $\theta$, $h\ \leq\ \theta$. As $h\ \rightarrow\ +\infty$, $F_2(h)$ tends to an horizontal positive asymptote, occurring at $(J\ \tau_{rec}^{-1}\ +\ I_{ext})$. For values of $h$ larger than $\theta$, the first derivative of $F_2(h)$ is always positive,  indicating that the function is monotonically increasing. In the same range of $h$, the second derivative is instead always negative, therefore indicating that the function is convex. Moreover, the tangent line to $F_2(h)$ at $h=\theta$ is the steepest of all the tangents to subsequent points ($h>\theta$).

\begin{equation}
  \left\{
    \begin{array}{l}
     \dot{F_2}(h)\  =\ \alpha\ A\ U\ / \left[ 1\ +\ \alpha\ \tau_{rec}\ U\ \left( h\ -\ \theta\right)\right]^2\\[3mm]
     \ddot{F_2}(h)\  =\ -\ 2\ \alpha^2\ \tau_{rec}\ A\ U^2\ / \left[ 1\ +\ \alpha\ \tau_{rec}\ U\ \left( h\ -\ \theta\right)\right]^3\\[3mm]
    \end{array}
  \right.
\label{graphical_solution2}
\end{equation}

The value of the slope of the tangent line at $F_2(h)$ at $h=\theta$ is particularly informative when drawing $F_2(h)$, since at any coordinates $h>\theta$ all the other tangent lines are by definition less steep than it. This maximal slope ($\dot{F_2}(\theta)\ =\ \alpha\ A\ U$) can then be used as a necessary condition for the intersections between $F_2(h)$ and the unitary slope line, at least when $I_{ext}\ \leq \theta$. 
%

When $I_{ext}\ \leq 0$, there is always an intersection between $F_2(h)$ and $F_1(h)$ at $h\ =\ I_{ext}$. This is a stable equilibrium point (\ie see below for the discussion of the stability). If $ \alpha\ A\ U\ \leq\ 1$, there will be for $I_{ext}\ \leq 0$ no other intersections, since $F_1(h)\ =\ h$ has unitary slope itself. Given the necessary condition $\alpha\ A\ U\ >\ 1$, there exist a minimal value for those parameters, above which other two intersections (\ie one stable and one unstable) with the unitary slope lines occur (compare \RefFig{fig:F2plot}A,B, and C, which were obtained for increasing values of $A$).

Determining analytically such values requires imposing the condition where the unitary slope line becomes tangent to $F_2(h)$. Mathematically this can be expressed by stating that when $I_{ext}\ \leq\ \theta$ there is a specific (intersection) point $h_0\ >\ \theta$ between $F_1(h)$ and $F_2(h)$. By definition, this point lays on the unitary slope line $F_2(h_0)\ =\ h_0$, with $\dot{F}_2(h_0)\ =\ 1$, \ie $F_2(h)$ has a unitary first derivative at $h_0$ (see \RefFig{fig:F2plot}B).

\begin{equation}
  \left\{
    \begin{array}{l}
     F_2(h_0)\  =\ \alpha\ A\ U\ (h_0 - \theta)\ / \left( 1\ +\ \alpha\ \tau_{rec}\ U\ (h_0 - \theta)\right)\ +\ I_{ext}\ =\ h_0\\[3mm]
     \dot{F_2}(h_0)\  =\ \alpha\ A\ U\ / \left[ 1\ +\ \alpha\ \tau_{rec}\ U\ \left( h_0\ -\ \theta\right)\right]^2\ =\ 1\\[3mm]
    \end{array}
  \right.
\label{critical_value}
\end{equation}

Simplifying the algebraic manipulations, we note the apparent similarities between the two equations above. We express (part of the) numerator and denominator of the right hand sides of each equations, by denoting the common terms as $N_0$ and $D_0$, respectively, as it follows

\begin{equation}
  \left\{
    \begin{array}{l}
     F_2(h_0)\  =\ N_0\  (h_0 - \theta)\ / D_0\ +\ I_{ext}\ =\ h_0\\[3mm]
     \dot{F_2}(h_0)\  =\ N_0\ /\ D_0^2\ =\ 1\\[3mm]
    \end{array}
  \right.
\label{critical_value2}
\end{equation}

Thus, from the second equation we derive $N_0\ /\ D_0\ =\ D_0$, and we substitute it in the first, obtaining

\begin{equation}
h_0\ =\ \theta\ +\ \sqrt{\left(\theta\ -\ I_{ext}\right)/\ \left( \alpha\ U\ \tau_{rec}\right)}
\end{equation}

The above expression is of course only defined when the argument of the square root is positive, which 
is consistent with our previous hypothesis (\ie $I_{ext}\ <\ \theta$). One can now replace $h_0$ in the second equation of \RefEqs{critical_value2} and obtain the corresponding critical value of $A$, associated to the existence of such a (double) intersection (\RefFig{fig:F2plot}B):

\begin{equation}
A_{0}\ =\ \left(1\ +\ \sqrt{\alpha\ U\ \tau_{rec}\ \left(\theta\ -\ I_{ext}\right)} \right)^2\ /\ \left( \alpha\ U\right)
\end{equation}

In summary, when $I_{ext}\ \leq\ \theta$ there is always one (stable) equilibrium at $h\ =\ I_{ext}$ and of possibly other two intersections (one stable and one unstable; compare \RefFig{fig:F2plot}A,B, and C), depending on the strength of $A$ with respect to $A_0$.
For $I_{ext}\ >\ \theta$, the situations changes and the scenario simplifies considerably, with only one (stable) intersection for any other choice of the parameters (see \RefFig{fig:F2plot}D).
 
The analysis of the stability of the equilibrium points conclude our discussion. Since the dynamical system described \RefEqs{depressing_mean_field} is non-linear, stability of equilibrium points must be related to the linearized system. The linearization is obtained for each equilibrium point by first-order Taylor expansion of \RefEqs{depressing_mean_field} around that point.
Let's compactly rewrite \RefEqs{depressing_mean_field} as

\begin{equation}
  \left\{
    \begin{array}{l}
     \dot{h}\ =\ G_1\left(h,x\right)\ =\ \left(-h\ +\ A\ U\ x\ E\ +\ I_{ext}\right)\ / \tau\\ [3mm]
     \dot{x}\ =\ G_2\left(h,x\right)\ =\ \left(1\ -\ x\right) / \tau_{rec}\ +\ U\ x\ E\\[3mm]
    \end{array}
  \right.
\label{depressing_mean_field_vector}
\end{equation}

The linearized system, around a generic equilibrium point $(h_0, x_0)$, is then

\begin{equation}
  \left\{
    \begin{array}{l}
     \dot{h}\ \approx\ G_1\left(h_0,x_0\right)\ +\ \partial G_1/\partial h|_{(h_0, x_0)}\ \left(h-h_0\right)\ +\ \partial G_1/\partial x|_{(h_0, x_0)}\ \left(x-x_0\right)\\ [3mm]
     \dot{x}\ \approx\ G_2\left(h_0,x_0\right)\ +\ \partial G_2/\partial h|_{(h_0, x_0)}\ \left(h-h_0\right)\ +\ \partial G_2/\partial x|_{(h_0, x_0)}\ \left(x-x_0\right)\\ [3mm]
    \end{array}
  \right.
\label{depressing_mean_field_linearized}
\end{equation}

Assessing the stability of the above system reduces to studying the {\em Jacobian} matrix $M$:

\begin{equation}
M(h_0, x_0) =
\left( {\begin{array}{cc}
 \partial G_1/\partial h & \partial G_1/\partial x  \\
 \partial G_2/\partial h & \partial G_2/\partial x  \\
 \end{array} } \right)|_{(h_0, x_0)}
\end{equation}

By definition, it is possible to make explicit the Jacobian matrix as

\begin{equation}
M(h_0, x_0) =
\left( {\begin{array}{cc}
 (-1+\alpha\ A\ U\ x_0)/\tau & (\alpha\ A\ U\ (h_0\ -\ \theta))/\tau  \\
 -\alpha\ U\ x_0 & -\tau_{rec}^{-1}\ -\ \alpha\ U\ (h_0\ -\ \theta)  \\
 \end{array} } \right)
\end{equation}

In particular, the real part of the two eigenvalues associated to $M(h_0, x_0)$ has been analyzed for each equilibrium point $(h_0, x_0)$. The eigenvalues $\lambda_{1,2}$ were computed as the roots of the following algebraic second order equation

\begin{equation}
det\left(I\ -\ \lambda\ M(h_0, x_0)\right)\ =\ 0
\end{equation}

\noindent where $I$ indicated the $2 \times 2$ identity matrix and $det()$ indicates the computation of the 
determinant of a square matrix. When at least one of the eigenvalue had positive real part, the equilibrium point was classified as {\em unstable}. When both eigenvalues had negative real parts, the equilibrium point was classified as {\em stable}. Nothing can be however concluded on the stability of the non-linear system, in the cases in which one or both eigenvalues have zero real part (and the other has negative real part). Stable and unstable equilibrium points have been graphically represented as circles and squares in \RefFig{fig:F2plot}, respectively, as well as in Figure 4C-E.

\subsection{A single population with non-depressing facilitating synapses}
 
The Eqs. 9-10 of the main text can be again simplified under the hypothesis of very fast recovery from depression $\tau_{rec}$. The mean-field firing rate dynamics of a single neuronal population, recurrently connected by short-term plastic synapses, can be rewritten as:

\begin{equation}
  \left\{
    \begin{array}{l}
     \tau\ \dot{h}\ =\ -h\ +\ A\ u\ E\ +\ I_{ext}\\ [3mm]
     \dot{u}\ =\ \left(U\ -\ u\right) / \tau_{facil}\ +\ U\ \left(1\ -\ u\right)\ E\\[3mm]
    \end{array}
  \right.
\label{facilitating_mean_field}
\end{equation}

While this hypothesis is not as realistic as the one of the previous section, it is preparatory for the analysis of the full model. As for the previous case, we consider the derivation of the equilibrium points for $h$ and $u$ as well as the assessment of their stability. Substituting the definitions of equilibrium,  $\dot{h}\ =\ 0$ and $\dot{u}\ =\ 0$, into \RefEqs{facilitating_mean_field},  we get

\begin{equation}
  \left\{
    \begin{array}{l}
     h =\ A\ u\ E\ +\ I_{ext}\\ [3mm]
     u\ =\ U\ \left(1\ +\ E\ \tau_{facil}\right)\ / \left( 1\ +\ U\ \tau_{facil}\ E\right)\\[3mm]
    \end{array}
  \right.
\label{facilitating_mean_field_steadystate}
\end{equation}

Using the second equation to replace $u$ as it appears in the first, one obtains an implicit equation in $h$:

\begin{equation}
    h =\ A\ U\ \left(1\ +\ E\ \tau_{facil}\right)\ E\ / \left( 1\ +\ U\ \tau_{facil}\ E\right)\ +\ I_{ext}
\label{phase_space2}
\end{equation}

As for \RefEq{phase_space}, numerical methods can be used for solving \RefEq{phase_space2}, looking for values of $h$ that satisfy the equivalence given a specific set of parameters $A$, $U$, $\tau_{facil}$, and $I_{ext}$. The solution(s) of \RefEq{phase_space2} can be also graphically interpreted as the intersection(s) in the cartesian plane of two functions of $h$, $F_1(h)$ and $F_3(h)$ as defined below

\begin{equation}
  \left\{
    \begin{array}{l}
     F_1(h)\ =\ h\\ [3mm]
     F_3(h)\  =\ A\ U\ \left(1\ +\ E\ \tau_{facil}\right)\ E\ / \left( 1\ +\ U\ \tau_{facil}\ E\right)\ +\ I_{ext}\\[3mm]
    \end{array}
  \right.
\label{graphical_solution3}
\end{equation}


We observe that by definition of $E\ =\ \left[\alpha\ (h - \theta)\right]_+$, $F_3(h)$ is zero when $h\ \leq\ \theta$. When $h\ \rightarrow\ +\infty$, the function diverges to infinity, but it can also be approximated by the straight line $F_3(h)\ \approx\ \alpha\ A\ h$. We also note that for values of $h$ larger than $\theta$, the first derivative of $F_3(h)$ is positive,  indicating that the function is monotonically increasing. In the same range of $h$, the second derivative is also positive, therefore indicating that the function is concave.

\begin{equation}
  \left\{
    \begin{array}{l}
     \dot{F_3}(h)\  =\ \frac{\alpha\ A\ U\ \left[1\ +\ 2\ \alpha\ \tau_{facil}\ (h\ -\ \theta)\ +\ \alpha^2\ \tau_{facil}^2\ U\ (h\ -\ \theta)^2\right]}{\left[ 1\ +\ \alpha\ \tau_{facil}\ U\ (h\ -\ \theta)\right]^2}\\[3mm]
     \ddot{F_3}(h)\  =\ 2\ \alpha^2\ \tau_{facil}\ A\ U\ (1\ -\ U)\ /\ \left[ 1\ +\ \alpha\ \tau_{facil}\ U\ (h\ -\ \theta)\right]^3\\[3mm]
    \end{array}
  \right.
\label{graphical_solution4}
\end{equation}

As opposed to the previous case, the value of the slope of the tangent line at $F_3(h)$ at $h\ =\ \theta$ is not  particularly relevant when drawing $F_3(h)$, since at any coordinates $h\ >\ \theta$ all the other tangent lines are by definition steeper than it. The minimal slope is $\dot{F_3}(\theta)\ =\ \alpha\ A\ U$ can then used in combination with the asymptotic approximation $F_3(h)\ \approx\ \alpha\ A\ h$ (\ie the maximal slope is $\alpha\ A$). It is then clear that, for $0\ \leq\ I_{ext}\ \leq \theta$, a sufficient and necessary conditions for having always two equilibrium points (\ie one stable at the value $h\ =\ I_{ext}$ when $0\ \leq\ I_{ext}\ \leq\ \theta$, and the other unstable) is represented by $\alpha\ A\ >\ 1$ (\RefFig{fig:F3plot}A,B). 
All the considerations on how to assess the dynamical stability of these equilibrium points hold, and the expression of the {\em Jacobian} matrix $M$, whose eigenvalues determine the stability, is given below:

\begin{equation}
M(h_0, u_0) =
\left( {\begin{array}{cc}
 (-1+\alpha\ A\ u_0)/\tau & (\alpha\ A\ U\ (h_0\ -\ \theta))/\tau  \\
 -\alpha\ U\ (1\ -\ u_0) & -\tau_{facil}^{-1}\ -\ \alpha\ U\ (h_0\ -\ \theta)  \\
 \end{array} } \right)
\end{equation}

\subsection{Single population with short-term plastic synapses}

In the general case, the mean-field equations of a single neuronal population, recurrently connected by short-term plastic synapses, are given by
\begin{equation}
  \left\{
    \begin{array}{l}
     \tau\ \dot{h}\ =\ -h\ +\ A\ u\ x\ E\ +\ I_{ext}\\ [3mm]
     \dot{x}\ =\ \left(1\ -\ x\right) / \tau_{rec}\ - \ u\ x\ E\\[3mm]
     \dot{u}\ =\ \left(U\ -\ u\right) / \tau_{facil}\ +\ U\ \left(1\ -\ u\right)\ E\\[3mm]
    \end{array}
  \right.
\label{full_mean_field}
\end{equation}

\noindent with the neuronal gain function chosen as a threshold-linear relationship between input (mean) current 
$h$ and output firing rate $E\ =\ \left[\alpha\ (h - \theta)\right]_+$ (see also Eqs. 9-10 of the main text). 
The analysis of this system, including its equilibrium points, has been already given elsewhere \cite{BarakTsodyks:2007}. Supporting the necessary condition on the symmetry breaking by long-term plasticities 
mentioned in the Results section of the main text, here we derive a simple observation on the analytical properties of these equilibrium points. According to the definition, we substitute $\dot{h}\ =\ 0$,  $\dot{x}\ =\ 0$, and $\dot{u}\ =\ 0$ into \RefEqs{full_mean_field},  and get

\begin{equation}
  \left\{
    \begin{array}{l}
     h =\ A\ u\ x\ E\ +\ I_{ext}\\ [3mm]
     x\ =\ 1\ / \left( 1\ +\ u\ \tau_{rec}\ E\right) \\[3mm] 
     u\ =\ U\ \left(1\ +\ E\ \tau_{facil}\right)\ / \left( 1\ +\ U\ \tau_{facil}\ E\right)\\[3mm]
    \end{array}
  \right.
\label{full_mean_field_steadystate}
\end{equation}

By appropriate substitutions of $x$ and of $u$ into the first equation, it is possible to express it as $h\ =\ F_4(E(h))$,  an implicit equation in $h$:

\begin{equation}
F_4(E)=\frac{AU \left( E^{-1} + \tau_{facil}\right)}{E^{-2}+E^{-1} U \tau_{facil}+E^{-1}U \tau_{rec}+  U \tau_{facil} \tau_{rec}}+I_{ext}.
\end{equation}

We observe that for $h\ \rightarrow\ +\infty$, $E(h)\ \rightarrow\ +\infty$ and $F_4(E) \ \rightarrow\  A\ \tau_{rec}^{-1}\ +\ I_{ext}$, implying the existence of an horizontal asymptote. This intuitively suggests that for any choice of the other parameters compatible with the existence of multiple equilibrium points (\ie intersections between $F_4(h)$ and the unitary slope line), the uppermost equilibrium point (\ie always stable) will  change its location proportionally to $A$ and to $\tau_{rec}^{-1}$, for the same choice of $I_{ext}$. Hence, a high value of $\tau_{rec}$, as in depressing synapses, will give a lower asymptote versus a low value, as in facilitating synapses.

Let's now consider two independent populations of excitatory neurons, one recurrently connected by short-term depressing synapses (\ie $\tau_{rec}\ >\ \tau_{facil}$) and one by short-term facilitating synapses (\ie $\tau_{facil}\ >\ \tau_{rec}$) and both receiving identical non-zero external inputs $I_{ext}$. As for the previous considerations on the horizontal asymptote of $F_4(E)$, for an appropriate choice of $A$  (\ie large enough to have multiple equilibrium points) or for any value of $I_{ext}\ >\ \theta$, the firing rate uppermost equilibrium point of the facilitating population will always be larger than the firing rate uppermost equilibrium point of the depressing population.
Together with the specific firing rate dependence of STDP, arising from the triplet-interactions, this consideration rules out that reciprocal motifs of short-term depressing synapses will outnumber unidirectional motifs of facilitating synapses. The stability analysis for the depressing and facilitating populations (with the parameters used in our simulations) is provided in the main text (see Fig. 3D-F).

Assessing the stability of the above system reduces to linearization of the system around the fixed points by the use of a Taylor expansion and the study of the so called {\em Jacobian} matrix $M(h_0, x_0, u_0)$ of the system, which for the sake of completeness, we report  below:

\begin{equation}
\left( {\begin{array}{ccc}
(-1+\alpha A u_0 x_0)/\tau & (\alpha A u_0 (h_0-\theta))/\tau & (\alpha A x_0 (h_0-\theta))/\tau \\
-\alpha u_0 x_0 & -\tau_{rec}^{-1}-\alpha u_0 (h_0-\theta) & -\alpha x_0 (h_0-\theta) \\
\alpha U (1-u_0) & 0 & -\tau_{facil}^{-1}-\alpha U (h_0-\theta) \\
 \end{array} } \right)
\end{equation}

\subsection{The impact of recurrent inhibition}
 

We extend the description of the system given in  \RefEqs{depressing_mean_field}, to the
case where recurrent inhibition is explicitly accounted for. The mean-field firing rate dynamics of two neuronal populations, one excitatory and one inhibitory, recurrently connected by short-term excitatory plastic synapses and by frequency-independent inhibitory synapses (as in \RefFig{fig:pop_ei}), can be rewritten as:

\begin{equation}
  \left\{
    \begin{array}{l}
     \tau_e\ \dot{h_e}\ =\ -h_e\ +\ A_{ee}\ U\ x\ E\ -\ A_{ei}\ I\ +\ I_{ext}\\ [3mm]
     \tau_i\ \dot{h_i}\ =\ -h_i\ +\ A_{ie}\ U\ x\ E\\ [3mm]
     \dot{x}\ =\ \left(1\ -\ x\right) / \tau_{rec}\ +\ U\ x\ E\\[3mm]
    \end{array}
  \right.
\label{depressing_mean_field_with_inhibition}
\end{equation}

\noindent with $E\ =\ \left[\alpha_e\ (h_e - \theta_e)\right]_+$ and $I\ =\ \left[\alpha_i\ (h_i - \theta_i)\right]_+$.
We consider here the derivation of the equilibrium points for $h_e$, $h_i$, and $x$, and we indicate how their stability was assessed. Substituting the conditions $\dot{h_e}\ =\ 0$, $\dot{h_i}\ =\ 0$, and $\dot{x}\ =\ 0$ in \RefEqs{depressing_mean_field_with_inhibition}, we obtain three non-linear algebraic equations

\begin{equation}
  \left\{
    \begin{array}{l}
     h_e =\ A_{ee}\ U\ x\ E\ -\ A_{ei}\ I\ +\ I_{ext}\\ [3mm]
     h_i =\ A_{ie}\ U\ x\ E\\ [3mm]
     x\ =\ 1\ / \left( 1\ +\ U\ \tau_{rec}\ E\right)\\[3mm]
    \end{array}
  \right.
\label{depressing_mean_field_with_inhibition_steadystate}
\end{equation}

Using the second and third equations to replace $x$ and $I$ in the first, we get an implicit equation in the unknown $h_e$:

\begin{equation}
    h_e =\ \frac{A_{ee}\ U\ E}{1\ +\ U\ \tau_{rec}\ E}\ -\ A_{ei}\ \left[\alpha_i\ \left(\frac{A_{ie}\ U\ E}{1\ +\ U\ \tau_{rec}\ E} - \theta_i \right)\right]_+\ +\ I_{ext}
\label{phase_space_with_inhibition}
\end{equation}

We make the assumption that the synaptic coupling from the excitatory to the inhibitory population  is sufficiently strong $A_{ie}\ >\ \tau_{rec}\ \theta_i$, so that short-term depression of that pathway does not prevent steady recruitment of inhibition at higher firing rates of the excitatory population.  We also assume for simplicity that recurrent excitation is also sufficiently strong so that $A_{ee}\ >\ \alpha_i\ A_{ei}\ A_{ie}$.
Under these hypotheses, when $E$ is below a certain value, \ie $E\ <\ \theta_i\ /\ \left[U\ \left(A_{ie}\ -\ \tau_{rec}\ \theta_i\right)\right]$, the 
above implicit equation coincides with \RefEq{phase_space} and it can be written as if inhibition was not present:

\begin{equation}
    h_e =\ A_{ee}\ U\ E\ / \left( 1\ +\ U\ \tau_{rec}\ E\right)\ +\ I_{ext}
\label{phase_space_with_inhibition2}
\end{equation}

Instead, above that critical value for $E$ (\ie and therefore for $h_e$), \RefEq{phase_space_with_inhibition} does not change formally, apart from its coefficients: 

\begin{equation}
    h_e =\ \hat{A}_{ee}\ U\ E\ / \left( 1\ +\ U\ \tau_{rec}\ E\right)\ +\ \hat{I}_{ext}
\label{phase_space_with_inhibition3}
\end{equation}

\noindent with $\hat{A}_{ee}\ =\ A_{ee}\ -\ \alpha_i\ A_{ei}\ A_{ie}$ and $\hat{I}_{ext}\ =\ I_{ext}\ +\ \alpha_i\ \theta_i\ A_{ei}$. It is easy to verify that $0\ <\ \hat{A}_{ee}\ <\ A_{ee}$ and $\hat{I}_{ext}\ >\ I_{ext}$.

The critical value for the recurrent inhibitory inputs to affect the excitatory population can be translated into a condition on $h_e$, \ie $h_e\ >\ \ \theta_e\ +\ \theta_i\ /\ \left[\alpha_e\ U\ \left(A_{ie}\ -\ \tau_{rec}\ \theta_i\right)\right]$. Under our previous hypothesis, such a critical value for $h_e$ is always larger than the threshold $\theta_e$ for the activation of the excitatory neurons. There will exist a range of activation for $h_e$ above $\theta_e$, where the impact of inhibition is negligible. For larger activation $h_e$, inhibition kicks by step-wise decreasing the parameter $\hat{A}_{ee}$.

%

All in all, the presence of recurrent inhibition in the system does not alter qualitatively the conclusions on the existence of equilibrium points of the mean field description. The statement on the separation of the uppermost equilibrium points, associated respectively to the short-term depressing and the short-term facilitating networks, remains true, since the horizontal asymptote shares the same indirect proportionality relationship with the time constant $\tau_{rec}$ of recovery from depression.

\subsection{Single population with mixed synapses}

We consider the special case of a homogeneous network of neurons, whose connections to distinct target postsynaptic neurons can be simultaneously short-term depressing and short-term facilitating. For the sake of simplicity and for distinguishing this case from the mixed populations studied in the main text (see Fig. 4), neurons are assumed to be indistinguishable from each other. However, every neuron has a certain probability $p_D$ to establish a short-term depressing connection with its postsynaptic target. The same neuron has probability $1\ -\ p_D$ to establish a short-term facilitating connection to another postsynaptic neurons. Under these simplifying hypotheses, and by definition of conditional expected value \cite{Papoulis:2002}, the mean-field firing rate dynamics of the neuronal population, recurrently connected by short-term plastic synapses, is

\begin{equation}
  \left\{
    \begin{array}{l}
     \tau\ \dot{h}\ =\ -h\ +\ A\ \left[ p_D\ u_D\ x_D\ +\ (1-p_D)\ u_F\ x_F\right] \ E\ +\ I_{ext}\\ [3mm]
     \dot{x_D}\ =\ \left(1\ -\ x_D\right) / \tau_{rec_D}\ - \ u_D\ x_D\ E\\[3mm]
     \dot{u_D}\ =\ \left(U_D\ -\ u_D\right) / \tau_{facil_D}\ +\ U_D\ \left(1\ -\ u_D\right)\ E\\[3mm]
     \dot{x_F}\ =\ \left(1\ -\ x_F\right) / \tau_{rec_F}\ - \ u_F\ x_F\ E\\[3mm]
     \dot{u_F}\ =\ \left(U_F\ -\ u_F\right) / \tau_{facil_F}\ +\ U_F\ \left(1\ -\ u_F\right)\ E\\[3mm]
    \end{array}
  \right.
\label{mixed_mean_field}
\end{equation}

The cases $p_D\ =\ 0$ and $p_D\ =\ 1$ have been already examined in the previous sections. For intermediate values of $p_D$, an evaluation of the equilibria of \RefEqs{mixed_mean_field} has been carried out numerically, resulting in a qualitatively similar behavior to the extreme cases, with the location of the (stable) equilibrium points to be intermediate between those of a short-term depressing neuronal network and those of a short-term facilitation neuronal network. As expected, for increasing values of $p_D$ the location of all equilibrium points of $E(h)$ (if any) decreases monotonically.

\subsection{Emergence of motifs when initial intra- and inter-population coupling are identical}
A specific initial configuration for the intra- (\ie $J_{FF}$, $J_{DD}$) and inter-population synaptic efficacies (\ie $J_{FD}$, $J_{DF}$) has been indicated in the main text as a necessary condition for subsequent ``symmetry-breaking" of the spontaneously emerging firing rates, in the depressing and the facilitating subpopulations (Fig. 4A,C). In this section, we relax such a condition and show how, by an appropriate external stimulation protocol, the same symmetry breaking could occur. As a consequence, the emerge of connectivity motifs is generally not impaired when synaptic couplings are set to identical values (\ie $J_{FF}$ = $J_{DD}$ = $J_{FD}$ = $J_{DF}$). Other protocols and conditions might lead to the same configuration and here we focus on the simplest.

We assume that the two subpopulations receive a common external input and an alternating pulsed stimulus component. As in a recurring traveling wave of external activity, each subpopulation is alternatively exposed to an pulsating input component, so that both facilitatory and depressing subpopulations are activated but never at same time.
Due to intrinsic subpopulation properties, determined by short-term synaptic plasticities and reviewed in the previous sections, and as a direct consequence of the associative character of long-term plasticity (Fig. 4B,D) discussed in the main text, this stimulation protocol leads to stronger intra-population synaptic coupling and weaker inter-population synaptic coupling (see Fig. \ref{fig:weightForm}). As discussed in the results of the main text, such a configuration is retained indefinitely, even in the absence of external alternating stimulation.

%

In order to understand why such a stimulation protocol succeeds in developing inter- and intra-population coupling asymmetries, we must examine the STDP parameters (\eg) in its mean field formulation (see Eq. 11 of the man text). Let's assume that the firing rate $E_D$ of the depressing subpopulation is larger than the firing rate  $E_F$ of the facilitating subpopulation (\ie say, $E_D\ =\ k\ E_F$, with $k\ >\ 1$). This configuration of firing rates is forced by the external input component, which alternates in time and across the subpopulations.
Under these conditions, the STDP would modify intra-population synaptic coupling $J_{DD}$ as follows:
\begin{equation}
\Delta J_{DD}= \ -A_2^-\ \tau_{o_1}\ E_D^2\ +\ A_3^+\ \tau_{r_1}\ \tau_{o_2}\ E_D^3
\end{equation}

\noindent while for the intra-population synaptic couplings, the STDP results into: 

\begin{equation}
\Delta J_{DF}= - \frac{1}{k}\ A_2^-\ \tau_{o_1}\ E_D^2\ +\frac{1}{k}\ A_3^+ \ \tau_{r_1}\ \tau_{o_2}\ E_D^3
\end{equation}

\begin{equation}
\Delta J_{FD}= - \frac{1}{k}\ A_2^-\ \tau_{o_1}\ E_D^2\ +\frac{1}{k^2}\ A_3^+\ \ \tau_{r_1}\ \tau_{o_2}\ E_D^3
\end{equation}

It is now easy to prove that $\Delta J_{DD}\ >\ \Delta J_{DF}\ >\ \Delta J_{FD}$, hence identical initial values for $J_{DD}$, $J_{DF}$, and $J_{FD}$ would slowly modified heterogeneously and leading to $J_{DD}\ >\ J_{DF}$ and to $J_{DD}\ >\ J_{FD}$.
Similar considerations can be repeated when the firing rate $E_F$ of the facilitating subpopulation is larger than the firing rate  $E_D$ of the depressing subpopulation (\ie say, $E_F\ =\ k\ E_D$, with $k\ >\ 1$), concluding that all in all that the stimulation protocol would shape synaptic efficacies as $J_{DD}\ >\ J_{DF}$, $J_{DD}\ >\ J_{FD}$, $J_{FF}\ >\ J_{DF}$, and $J_{FF}\ >\ J_{FD}$, therefore privileging intra-population coupling to inter-population ones, as shown in Fig. \ref{fig:weightForm}B.

\subsection{Statistics of the symmetry index}

In order to quantify and describe concisely the symmetries of the emerging network connectivity matrix $\left[A_{ij} \right]$ of size $N\ \times\ N$, we defined the following quantity (but see \eg \cite{HubertBaker:1979}, for alternative definitions):

\begin{equation}
	s\ =\ 1\ -\ \frac{1}{\left(0.5\ N \left(N-1\right)\ -\ M\right)}\ \sum_{i=1}^{N}\sum_{j=i+1}^{N}{|A^*_{ij}\ -\ A^*_{ji}|}
\label{symmetry_measure}
\end{equation}

$s$ intuitively represents the mean absolute difference between elements that are on symmetric positions, with respect to the diagonal of the matrix. By definition, the elements $A^*_{ij}$ are obtained from $A_{ij}$ upon first normalizing its numerical values to the maximal allowed $A_{max}$ and then clipping them to a lower fraction $z$. For instance, choosing $z\ =\ 2/3$, if $A_{ij}\ >\ 2/3\ A_{max}$ then $A^*_{ij}\ =\ A_{ij}\ /\ A_{max}$, and otherwise $A^*_{ij}\ =\ 0$. In the \RefEq{symmetry_measure}, $M$ represents the number of null pairs $\{A^*_{ij},\ A^*_{ji}\}\ =\ \{0,\ 0\}$ as a consequence of clipping.
Then, $s$ can be rewritten in terms of an arithmetic average of a set of $K$ observations of a random variable $q$:

\begin{equation}
	s\ =\ 1\ -\ \frac{1}{K}\ \sum_{i=1}^{K}{q_i}
\label{symmetry_measure2}
\end{equation}
 
 Assuming for simplicity that each element of $\left[A_{ij} \right]$ is independently drawn from a uniform distribution (\ie between $0$ and $A_{max}$), the probability density distribution of $q$, $f_q(Q)$, can be derived analytically \cite{Papoulis:2002}. 
Because the arithmetic average is an unbiased estimator of the expected value of the random variable it samples (\ie in this case $q$) \cite{Papoulis:2002}, most of the statistical properties of $s$ can be immediately derived from the distribution of $q$. First of all, the expected value of $s$ is given by the expression below

\begin{equation}
	\avg{s}\ =\ A_{max}\ \left(1\ -\ \frac{1-z}{1-z^2}\ \left[ \frac{(1-z)^2}{3}\ +\ z\ (1+z)\right]\right)
\label{symmetry_measure_mean}
\end{equation}

Deriving the expression for the variance is less straightforward and requires estimating the expected value of $1/K$. In fact $K$ coincides with the number of terms in the double sum of \RefEq{symmetry_measure} and it is by definition not a fixed quantity, but a realization of a binomial random variable. An approximated expression for the variance of $s$ is then

\begin{equation}
	Var\{s\}\ \approx\ A_{max}^2\ \frac{2}{N\ (N-1)\ (1-z^2)}\ \left(1\ +\ \frac{2\ z^2}{N\ (N-1)\ (1-z^2)} \right)\ Var\{q\}
\label{symmetry_measure_var}
\end{equation}

where

\begin{equation}
	Var\{q\}\ =\ \frac{1}{1-z^2}\ \left( \frac{(1-z)^4}{6}\ +\ \frac{2\ z\ (1-z^3)}{3}\right)\ -\ (1-\avg{s})^2
\label{symmetry_measure_varq}
\end{equation}

The validity of all the above expressions have been tested and validated numerically, directly estimating the average and variance of $s$ across thousands of uniform random matrices $\left[A_{ij}\right]$, for several values of $z$ in the range $[0.1;0.9]$, finding an excellent agreement. 

Finally, from the Central Limit Theorem \cite{Papoulis:2002}, we can expect the density distribution of $s$ to be approximately Gaussian, at least for small values of $z$. 
By the above statistical expressions, when studying the impact of short- and long-term synaptic plasticity in shaping microcircuit connectivity motifs (see Figs. 1-2,5), we expressed the significance of the observed  values of $s$ as the chance level, \ie the (Gauss-distributed) probability that the observed value of $s$ could be obtained by chance from a random uniform matrix.

\subsection{Alternative STDP models}
{\subsubsection{Pair-based STDP model}}

By appropriately choosing $A_2^+$, $A_2^-$, and setting to zero both $A_3^+$ and $A_3^-$, earlier phenomenological models of pair-based STDP can be rephrased as a special case of the triplet-based model. For {the pair-based STDP}, each neuron of the network needs only two indicator variables, \ie $q_1$ and $o_1$, instead of four. In the lack of any firing activity of the $j$-th neuron, those variables exponentially relax to zero:
\begin{equation}
     \tau_{q_1}\ \dot{q}_{1_j}\ =\ -q_{1_j}\ \ \ \ \ \ \ \ \ \ 
     \tau_{o_1}\ \dot{o}_{1_j}\ =\ -o_{1_j}
\label{newSTDPindicators}
\end{equation}

As the $j$-th neuron fires, the variables must be instantaneously updated. For such update rule, there are two distinct scenarios determining how successive pre-post or post-pre events interact and affect synaptic efficacy: i) {\em all-to-all} spike pairs interactions, 

\begin{equation}
 q_{1_j}\ \rightarrow\ q_{1_j}\ +\ 1\ \ \ \ \ \
 o_{1_j}\ \rightarrow\ o_{1_j}\ +\ 1
\label{newSTDPindicators2}
\end{equation}

\noindent and ii) {\em nearest-spike} interactions, 

\begin{equation}
 q_{1_j}\ \rightarrow\ 1\ \ \ \ \ \
 o_{1_j}\ \rightarrow\ 1.
\label{newSTDPindicators3}
\end{equation}

\noindent where the update rules do not allow accumulation of effects. Finally, when the $j$-th neuron spikes, the following updates are performed over all the indexes $i$:

\begin{equation}
  \left\{
    \begin{array}{l}
     W_{ij}\ \rightarrow\ W_{ij}\ -\eta\ A_2^-\ o_{1_i}(t)\\
     W_{ji}\ \rightarrow\ W_{ji}\ +\eta\ A_2^+\ q_{1_i}(t)\\
      \end{array}
      \right.
 \label{newSTDPrule}
\end{equation}

Following the considerations by Pfister and Gerstner \cite{PfisterGerstnerNIPS2006}, pair-based models of STDP with all-to-all interactions must be excluded, as they do not reproduce realistic (\ie BCM) features of synaptic plasticity.
We then consider the triplet-based STDP model and alter some of its parameters as it follows: $A_3^+\ =\  A_3^- \ =\ 0$, $A_2^+\ =\ 4.5\ 10^{-3}, A_2^-\ =\ 7.1\ 10^{-3}$. We also replaced the {\em all-to-all} spike pairs interactions by a {\em nearest spike} interaction, by modifying the update rule for $q_1$ and $r_1$ (and for $q_2$ and $r_2$, although those state variables are anyway irrelevant, upon setting $A_3^+\ =\  A_3^- \ =\ 0$). By doing so, we obtain the {pair-based STDP} plasticity rule {matching the exact same temporal window of the STDP triplet model employed here}. In order to prove that there is correspondence in terms of the temporal window, but altered frequency-dependence, we subjected both the triplet-based and the pair-based models to 75 pairing events, at low frequency $10 Hz$. The STDP temporal windows at such a frequency are undistinguishable from each other (compare Fig. 2A with Fig. 3A). The frequency-dependence is computed across the same number of pairing events, imposing a pre-post or post-pre delay of $10 msec$. Note that for sufficiently large frequency of the pre-post pairs, the curves corresponding to pre-post and to post-pre become symmetric with respect to a horizontal line (see Fig. 3B; \ie around $75-80\%$, corresponding to long-term depression). Such a symmetry implies that in the case of random occurrence of pre-post or post-pre timing, the LTP and the LTD components would cancel each other on the average.

\subsubsection{Triplet-based anti-STDP model}

This model is obtained from the triplet-based, by setting $A_3^+\ =\  7.1\ 10^{-3}$, $A_3^- \ =\ 6.1\ 10^{-3}$, $A_2^+\ =\ 0$, and $A_2^-\ =\ 3.5\ 10^{-3}$, by leaving unchanged the update rules for $q_1$, $q_2$, $o_1$, and $o_2$ as in the original triplet model, and by modifying the actual weight update equations as it follows: when the $j$-th neuron spikes, the following updates are performed for all the indexes $i$:

\begin{equation}
  \left\{
    \begin{array}{l}
     W_{ij}\ \rightarrow\ W_{ij}\ +\eta\ \ o_{1_i}(t)\ \left[ A_2^+\ +\ A_3^+\ q_{2_j}(t-\epsilon)\right]\\
     W_{ji}\ \rightarrow\ W_{ji}\ -\eta\ \ q_{1_i}(t)\ \left[ A_2^-\ +\ A_3^-\ o_{2_j}(t-\epsilon)\right]\\
      \end{array}
      \right.
 \label{antiSTDPrule}
\end{equation}

Note that this is only a tentative proposal for an anti-STDP rule, since experimental of data is not yet available for all induction protocols earlier employed for STDP. In particular, we ignore the frequency-dependency of the anti-STDP and by the new parameter set we roughly leave it untouched (compare Fig. 2B with Fig. 3D).

\bibliography{biblio}

\newpage
\section*{Figures and Figure Legends}

\begin{figure}[htb!]
\centering
\includegraphics[scale=0.4]{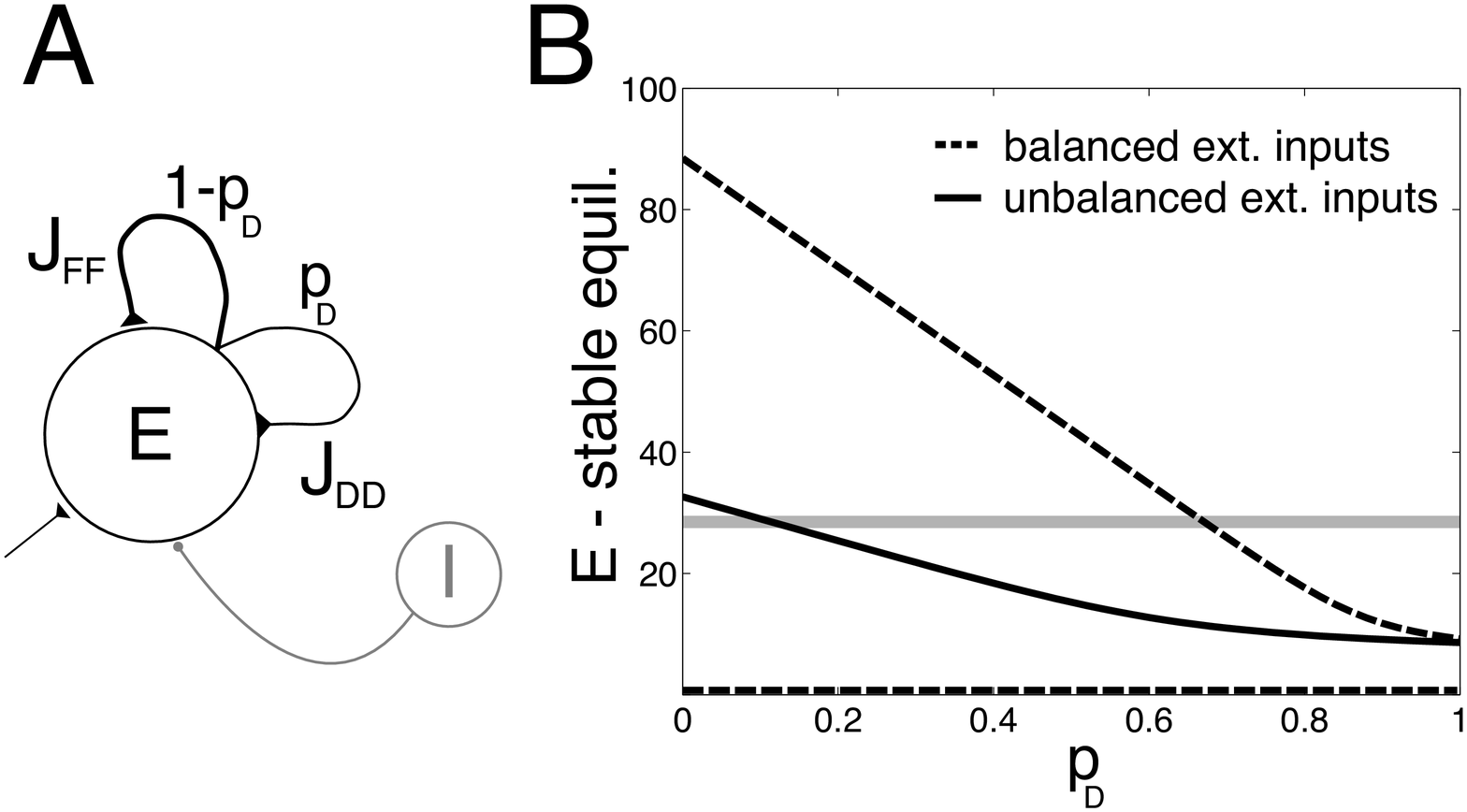}
\caption{{\bf Mean-field analysis of firing rate equilibria, in homogeneous networks with overlapping short-term synaptic properties and no long-term plasticity}.  Panel {\bf A} represents  the sketch of a recurrent network where a clear segregation between subpopulations of depressing- or facilitating-only synapses does not occur. A neuron has a probability $p_D$ of connecting to its postsynaptic target by a depressing synapse, and $1-p_D$ of connecting to its target by a facilitating synapse.  Panel {\bf B} plots the location of the equilibria of the firing rate $E$, under distinct external inputs conditions and for increasing values of $p_D$.  For the same parameters of \RefFig{fig:MeanField}, stable equilibria move as a function of $p_D$, taking intermediate values between the two extreme cases, \ie $p_D\ =\ 0$ and $p_D\ =\ 1$; compare to panels {\bf D-F} of \RefFigs{fig:MeanField}.
\label{fig:MixedMeanField}} 
\end{figure}

\begin{figure}[htb!]
\centering
\includegraphics[scale=0.46]{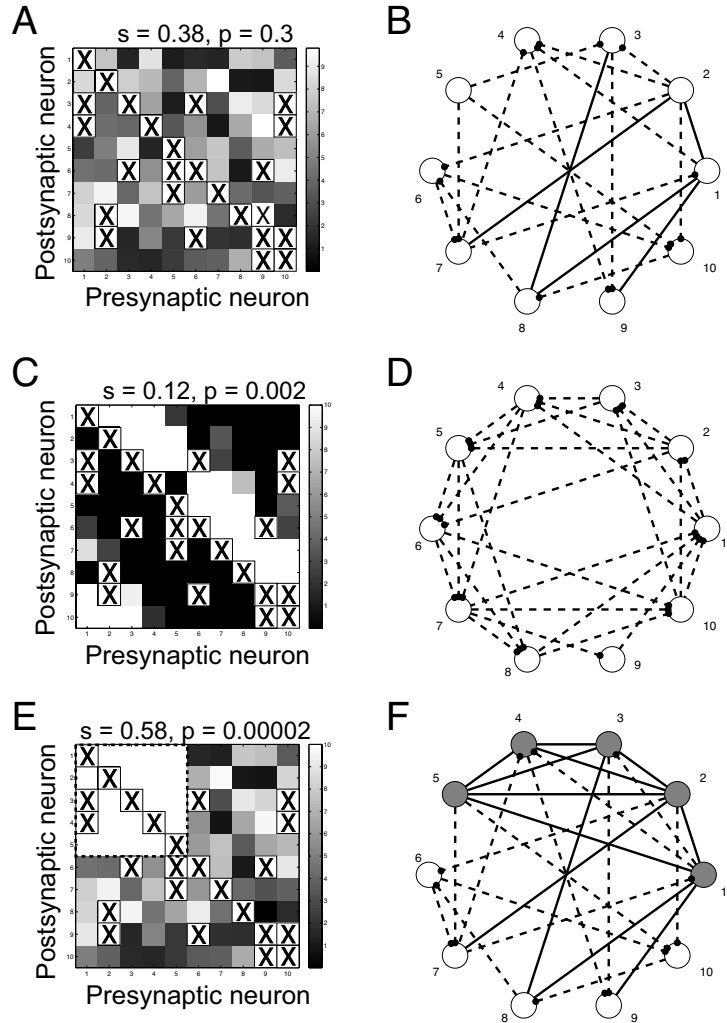}
{\caption{{\bf External activity may still induce reciprocal motifs emergence in depressing networks}. As in Figure 1, the synaptic connectivity matrix of a network of ten neurons was randomly initialized and pruned ({\bf A,B}, \ie pruning is indicated by the ``X" symbols). Internally generated activity, as in Figures 1-2, contributes to the emergence of non-random unidirectional motifs, resulting in an asymmetric matrix W ({\bf C,D}). 
However, if five units of the same network (gray circles) are externally stimulated above the STDP {\em critical} frequency (see the Results of the main text), a non-random connectivity emerges, featuring reciprocal motifs and  a symmetric connectivity submatrix ({\bf E,F}; upper left corner, dashed rectangle). The values indicated above panels {\bf A,C,E} represent the symmetry index and its significance.\label{fig:asinClopath}} }
\end{figure}

\begin{table*}[htb!]
\centering
\begin{tabular}{llll}
\hline
Symbol & Description & Value \\
\hline
$dt$ & Forward Euler method integration time step & $0.1\ msec$\\
$N$ & Number of simulated neurons & $10\ -\ 1000$\\
$c_m$ & Membrane capacitance & $281\ pF$\\
$g_{leak}$ & Membrane leak conductance & $30\ nS$\\
$E_{leak}$ & Resting membrane potential & $-70.6\ mV$\\
$E_{reset}$ & After-spike reset potential & $-70.6\ mV$\\
$\Delta_T$ & Spike steepness of the exponential IF model & $2\ mV$\\
$V_{\theta}$ & Spike emission threshold of the exponential IF model & $20\ mV$\\
$V_T$ & Threshold voltage parameter of the exponential IF model & $-50.4\ mV$\\
$\tau_{arp}$ & Absolute refractory period & $2\ msec$\\
$a$ & Voltage dependence coefficient of the spike frequency adaptation & $4\ nS$\\
$\Delta_x$ & Spike-timing dependence parameter of the spike frequency adaptation & $0.0805\ nA$\\
$\tau_{x}$ & Time constant of the spike frequency adaptation & $144\ msec$\\
$\tau_{syn}$ & Excitatory postsynaptic currents decay time constant & $5\ msec$\\
$U_D$ & Release probability, for depressing synapses & $0.8$\\
$U_F$ & Release probability, for facilitatory synapses & $0.1$\\
$\tau_{rec\ D}$ & Time constant of recovery from depression, for {\em depressing} synapses & $900\ msec$\\
$\tau_{rec\ F}$ & Time constant of recovery from depression, for {\em facilitating} synapses & $100\ msec$\\
$\tau_{facil\ D}$ & Time constant of recovery from facilitation, for {\em depressing} synapses & $100\ msec$\\
$\tau_{facil\ F}$ & Time constant of recovery from facilitation, for {\em facilitating} synapses & $900\ msec$\\
$A_2^-$ & STDP model LTD amplitude for post-pre event & $7.1\ 10^{-3}$\\
$A_3^-$ & STDP model LTD amplitude for post-pre event (triplet-term) & $0$\\
$A_2^+$ & STDP model LTP amplitude for pre-post event & $0$\\
$A_3^+$ & STDP model LTP amplitude for pre-post event (triplet-term) & $6.5\ 10^{-3}$\\
$\tau_{q_1}$ & STDP model decay time of presynaptic indicator $q_1$& $16.8\ msec$\\
$\tau_{q_2}$ & STDP model decay time of presynaptic indicator $q_2$ & $101\ msec$\\
$\tau_{o_1}$ & STDP model decay time of postsynaptic indicator $o_1$ & $33.7\ msec$\\
$\tau_{o_2}$ & STDP model decay time of postsynaptic indicator $o_2$ & $114\ msec$\\
$A_{i\ j}$ & Maximal synaptic efficacy & $6-12\ pA$\\
$W_{max}$ & Upper boundary for STDP dimensionless scaling factor $W_{ij}$ & $5$\\
$\theta$ & Threshold of the frequency-current response curve for mean-field models & $3$\\
$\eta$ & STDP plasticity rate & $1$\\
\hline
\end{tabular}
\caption{Parameters employed in the simulations: STDP parameters are as in the {\em minimal all-to-all} triplet model described in Pfister and Gerstner (2006); short-term depression and facilitation parameters as in (Wang {\em et al.}, 2006); neuron parameters are as in (Clopath {\em et al.}, 2010).}
 \label{tab:params}
\end{table*}

\renewcommand{\thefigure}{S\arabic{figure}}
\setcounter{figure}{0}


\newpage

\begin{figure}[htb!]
\centering
\includegraphics[scale=0.5]{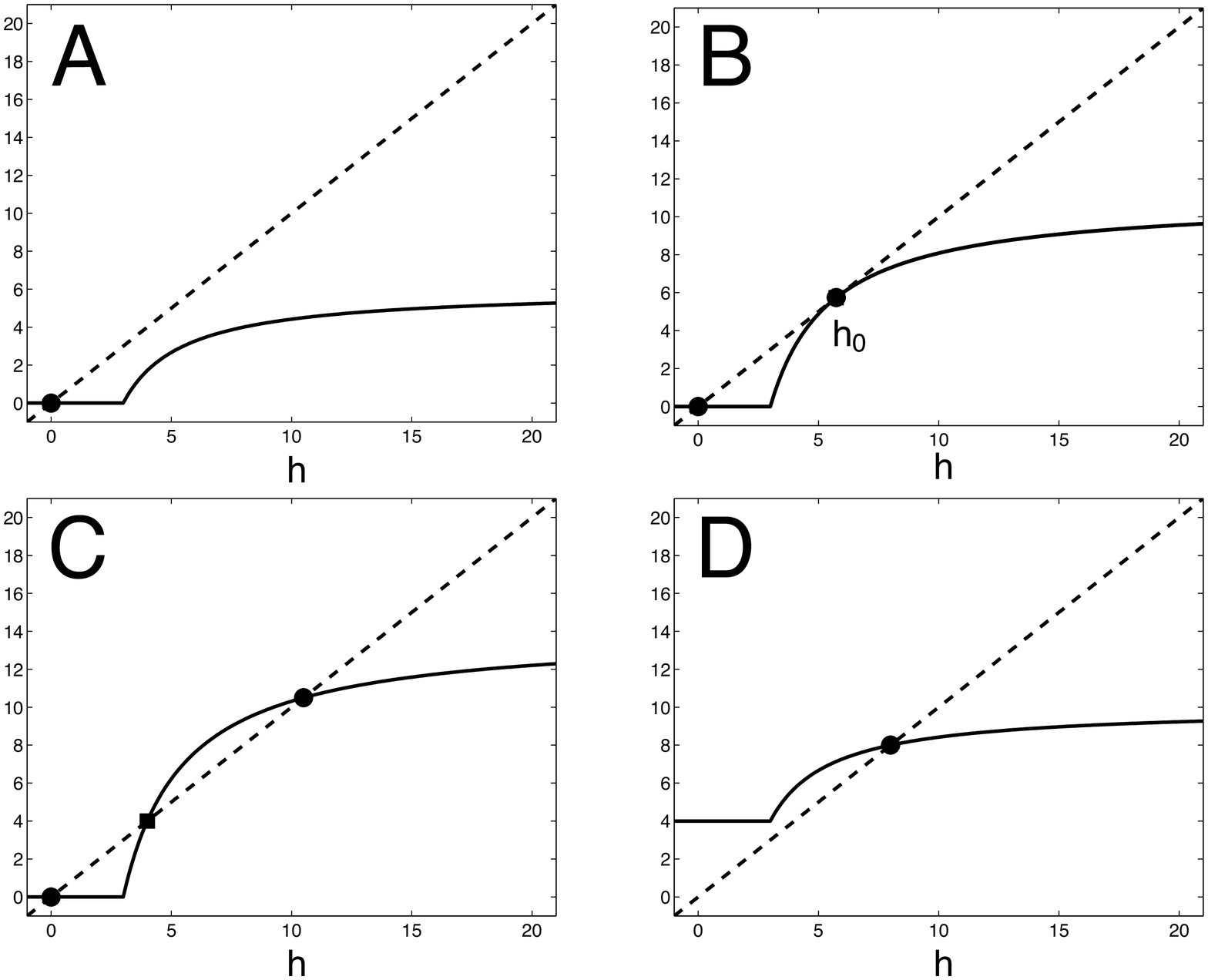}
\caption{{\bf Graphical representation of the function $F_2(h)$, for different values of $A$ and $I_{ext}$}. 
The function $F_2(h)$ has been plotted for different values of the maximal synaptic efficacy $A$ and external input $I_{ext}$, resulting in one intersection only ({\bf A} - $A\ =\ 3$, $I_{ext}\ =\ 0$) , three intersections with two of them coincident with each other ({\bf B} - $A\ =\ 5.488$, $I_{ext}\ =\ 0$), three distinct intersections ({\bf C} - $A\ =\ 7$, $I_{ext}\ =\ 0$), and finally one intersection for larger external input ({\bf D} - $A\ =\ 3$, $I_{ext}\ =\ 4$). The remaining parameters were: $U\ =\ 0.8$, $\tau_{rec}\ =\ 500\ msec$, $\theta\ =\ 3$, $\alpha\ =\ 1$. The scripts to generate these plots and to carry out asymptotic analysis on the stability of the equilibrium points, see the text, are available online from the ModelDB (accession number 143082).}
\label{fig:F2plot}
\end{figure}

\begin{figure}[htb!]
\centering
\includegraphics[scale=0.5]{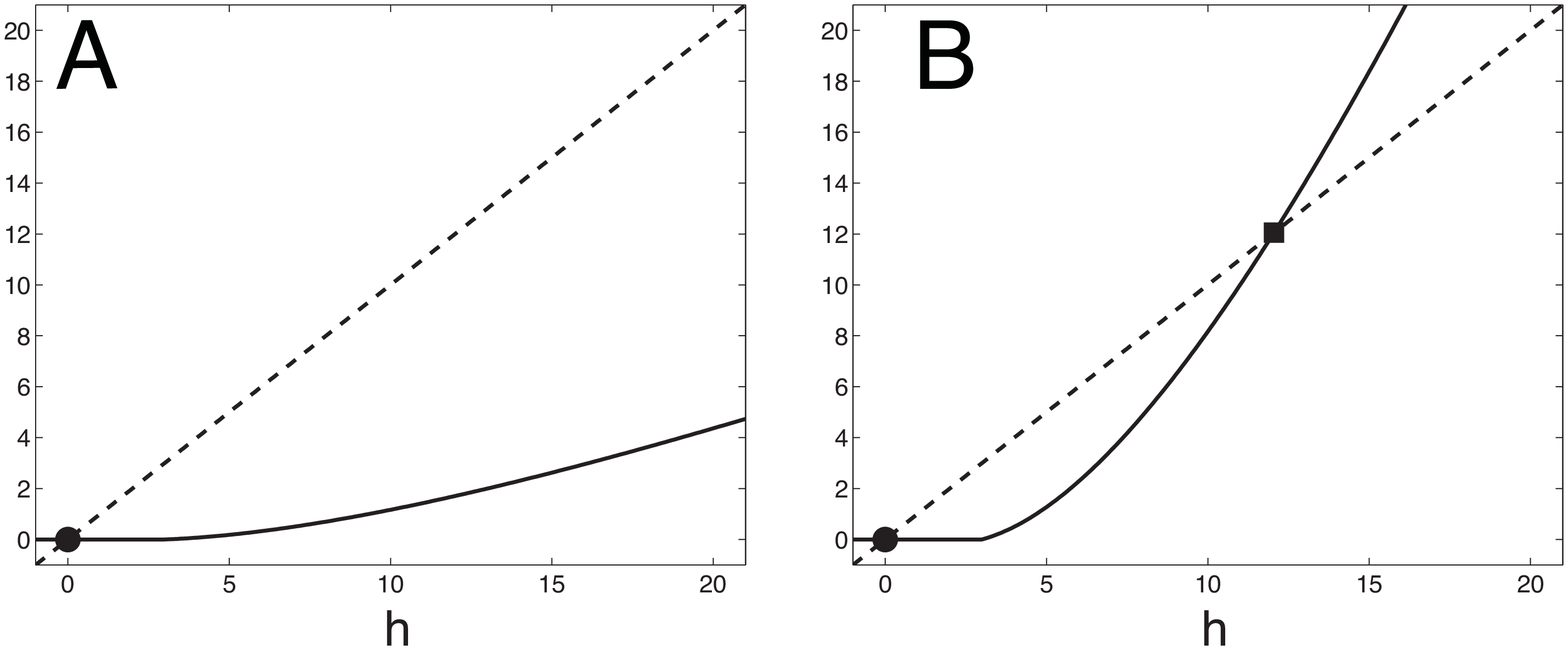}
\caption{{\bf Graphical representation of the function $F_3(h)$, for different values of $A$}. 
The function $F_3(h)$ has been plotted for different values of the maximal synaptic efficacy $A$ and external input $I_{ext}$, resulting in one ({\bf A} - $A\ =\ 0.5$, $I_{ext}\ =\ 0$) or two intersections ({\bf B} - $A\ =\ 3.5$, $I_{ext}\ =\ 0$). The remaining parameters were: $U\ =\ 0.1$, $\tau_{facil}\ =\ 500\ msec$, $\theta\ =\ 3$, $\alpha\ =\ 1$. The scripts to generate these plots and to carry out asymptotic analysis on the stability of the equilibrium points, see the text, are available online at the ModelDB (accession number 143082).}
\label{fig:F3plot}
\end{figure}
\begin{figure}[htb!]
\centering
\includegraphics[scale=0.2]{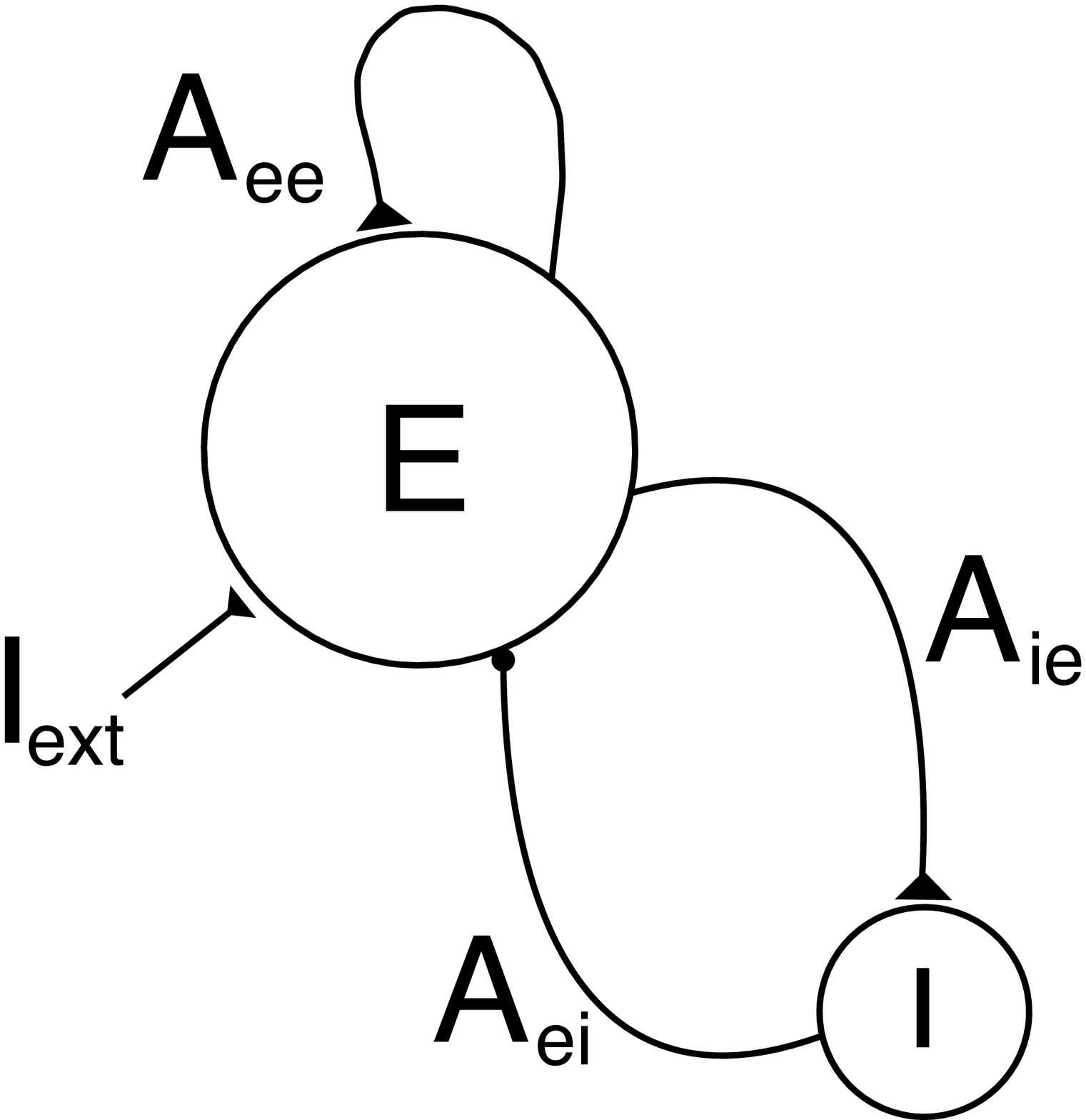}
\caption{{\bf Mean field description for two recurrently connected populations of excitatory and inhibitory neurons}. 
Excitatory neurons are recurrently connected by short-term plastic synapses and project to a population of inhibitory neurons. Inhibitory neurons project back to the excitatory population with non-plastic, linear synapses.}
\label{fig:pop_ei}
\end{figure}
\begin{figure}[htb!]
\centering
\includegraphics[scale=0.4]{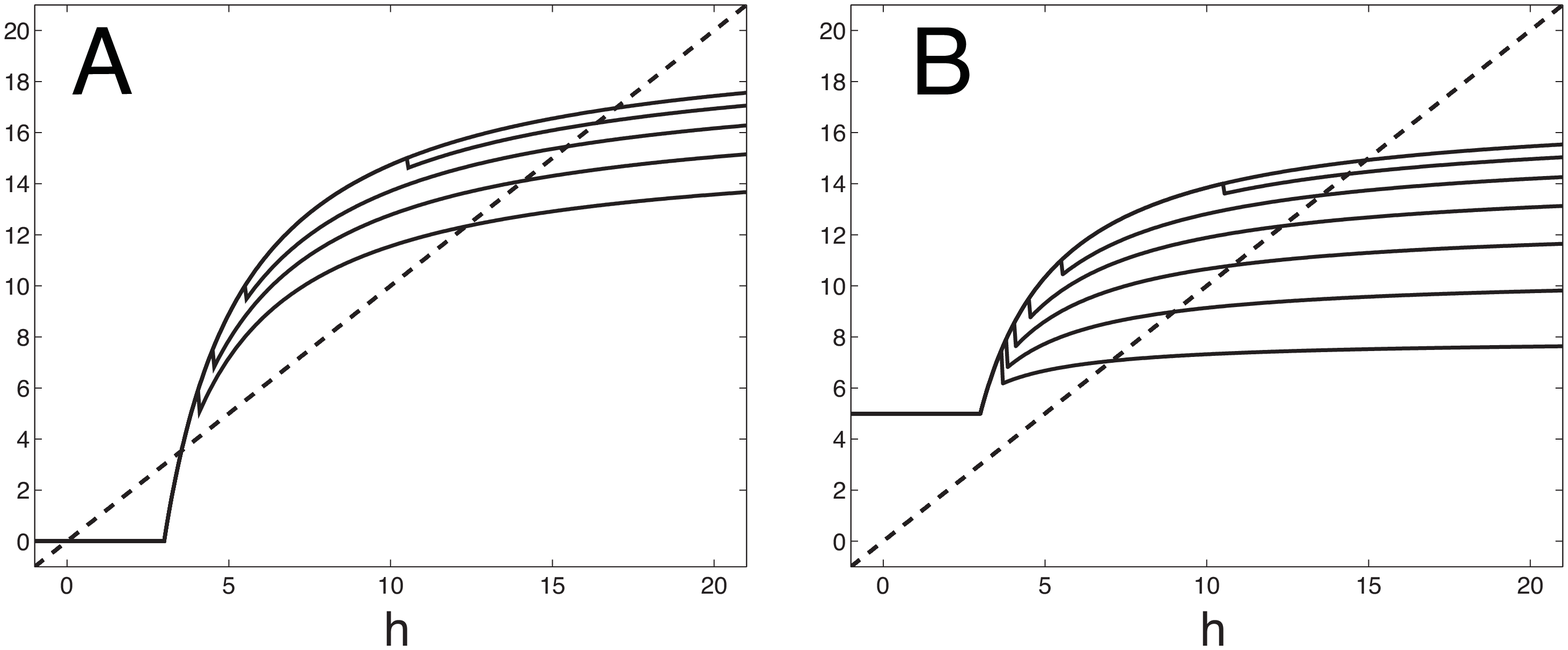}
\caption{{\bf Graphical representation of the equivalent of functions $F_2(h)$ and $F_3(h)$, in the presence of inhibition}. 
Increasing the coupling between the excitatory and the inhibitory population substantially bends down the curves previously analyzed as $F_2(h)$ and $F_4(h)$, lowering the firing rate of the equilibrium points ($A_{ei}\ =\ A_{ie}\ /\ 10$; {\bf A} - $I_{ext}\ =\ 0$, {\bf B} - $I_{ext}\ =\ 5$).}
\label{fig:pop_ei2}
\end{figure}
\begin{figure}[htb!]
\centering
\includegraphics[scale=0.62]{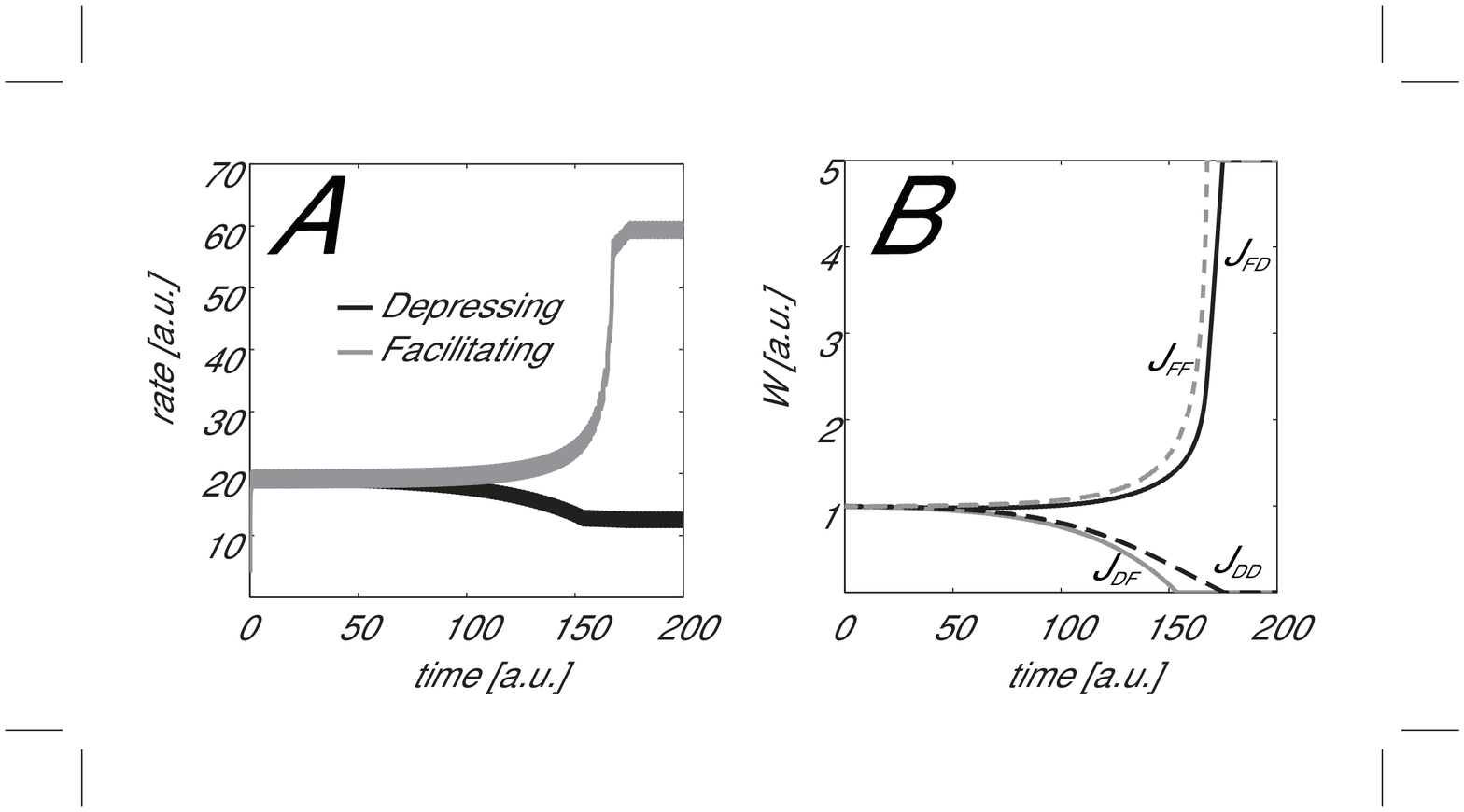}
\caption{{\bf Formation of asymmetric intra and extra population synapses}. 
The subopulations $D$ and $F$ of Fig. 4A receive a common input current ($I_C\ =\ 12.5$) and an alternating stimulation component ($I_{ED}$, $I_{EF}$), oscillating periodically between two amplitude levels (\ie 2 and 0) every 1 [a.u.] of time: whenever $I_{ED}\ =\ 0$, $I_{EF}\ =\ 2$, and vice versa. The temporal evolution of the firing rate of each population is shown in {\bf A}, while the corresponding long-lasting plastic changes of the synaptic coupling (\ie $J_{FF}$, $J_{DD}$, $J_{FD}$, $J_{DF}$) is plotted in {\bf B}, upon initialization to the same value (\ie $1$).
The same final configuration is generally obtained even by randomly initializing $J_{FF}$, $J_{DD}$, $J_{FD}$, and $J_{DF}$ by a Gaussian distribution with mean $1$ and standard deviation $0.001$, for 90 out of 100 simulation runs, demonstrating a degree of robustness.}
\label{fig:weightForm}
\end{figure}

\begin{figure}[htb!]
\centering
\includegraphics[scale=0.5]{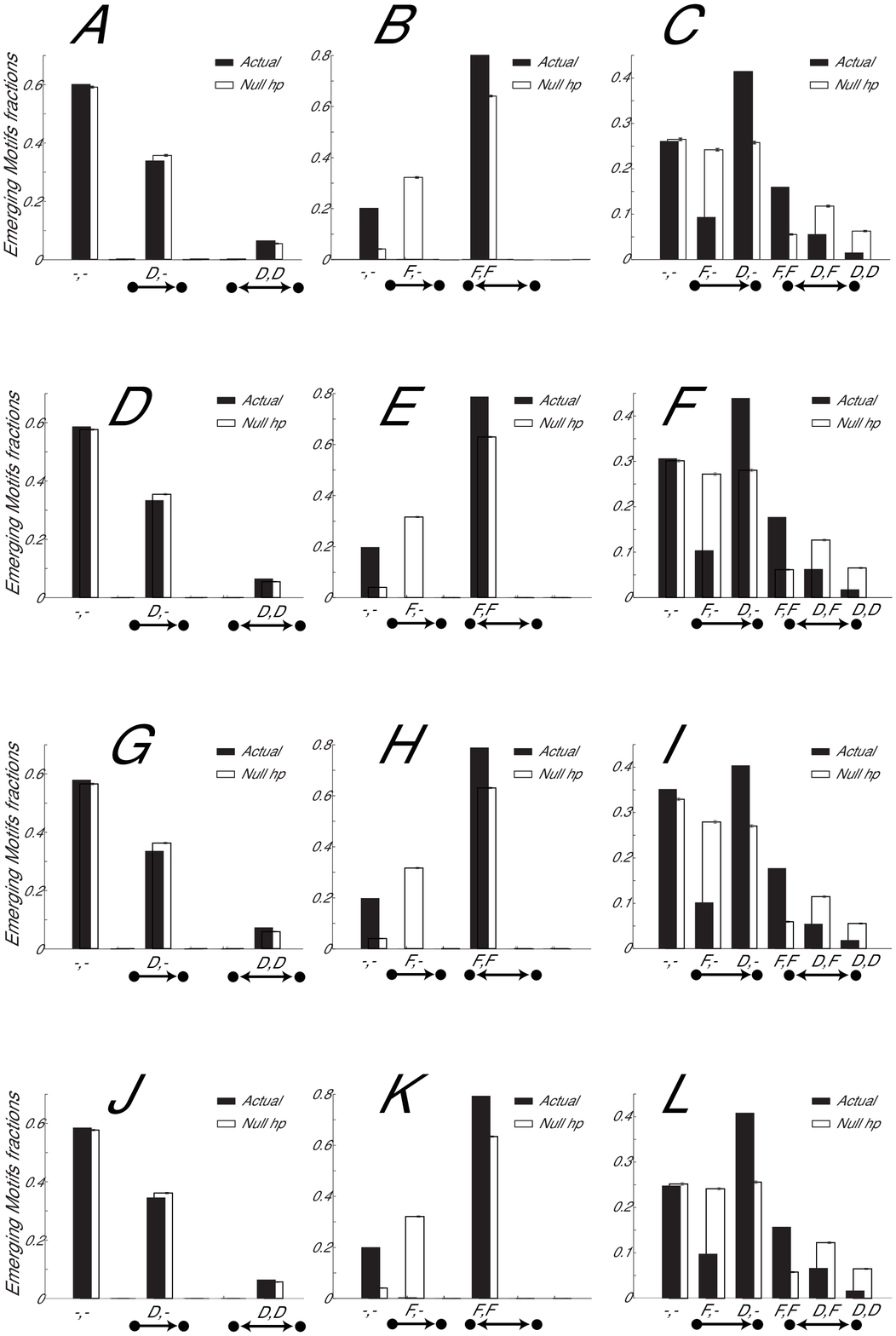}
\caption{{\bf External stimulation protocols and models details do not affect motifs emergence}. The figure examines the fraction of motifs, emerging when neurons receive periodic input wave-like stimulation ({\bf A,B,C}, exactly as in Figure 6A-C, for ease of comparison). However, when the periodic stimulation is omitted ({\bf D,E,F}), as well as when each neuron receives instead a ten percent shared random background inputs with each other ({\bf G,H,I}), very similar results emerge. The simulations of panels {\bf D,E,F}, when repeated without spike-frequency adaptation mechanisms from each unit of the network(s) ({\bf J,K,L}), still give rise to the same results. \label{fig:newLargeNetworks}} 
\end{figure}
\begin{figure}[htb!]
\centering
\includegraphics[scale=0.5]{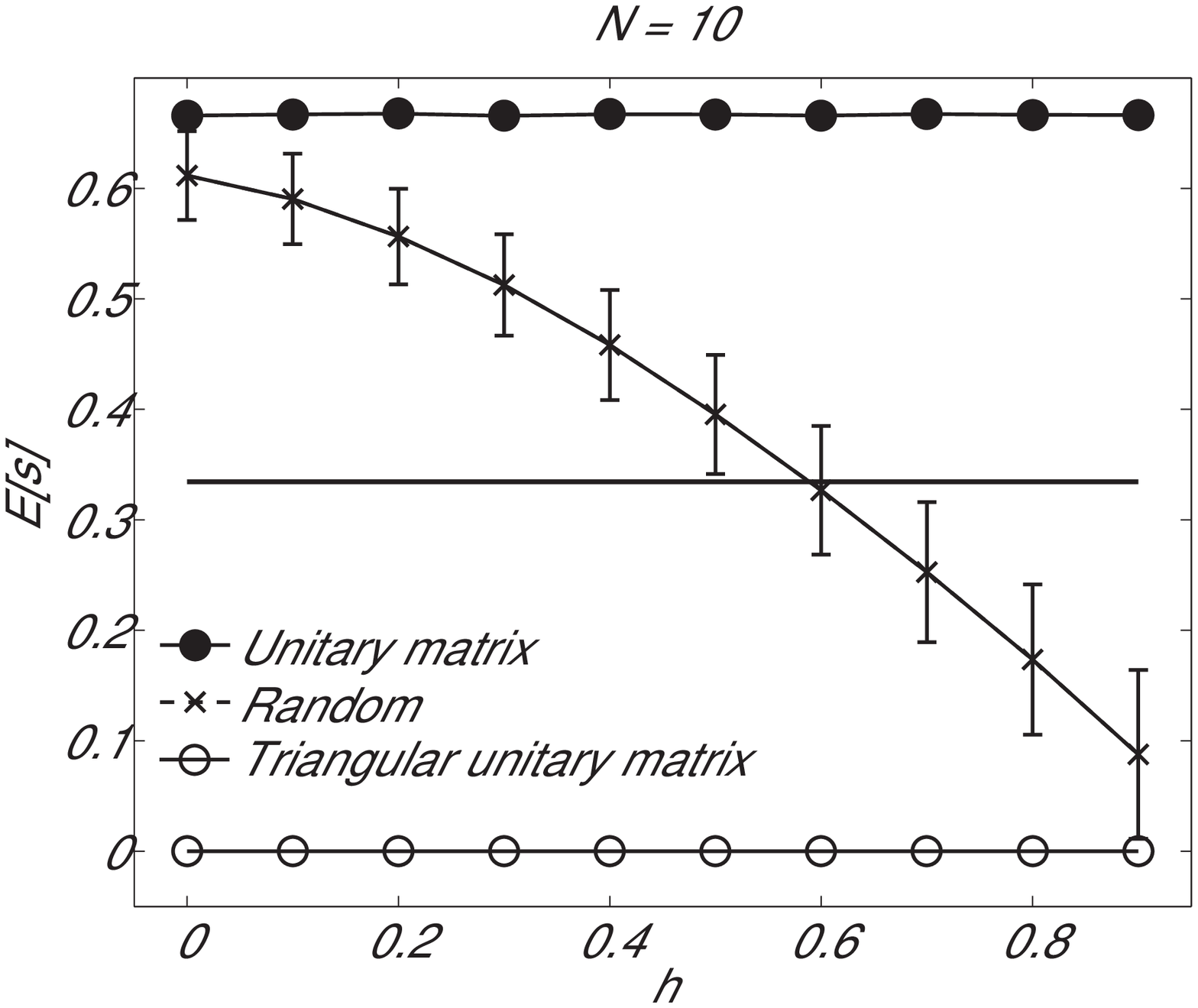}
{\caption{{\bf  Impact of the clipping threshold parameter $h$ on the symmetry measure $s$}. The continuous trace and $\times$ markers show the expectation of the symmetry index for a random connectivity matrix with $20\%$ of its elements randomly pruned, over $10000$ samples of $10\ \times\ 10$ matrices, with error bars indicating the standard deviation. The simulations were repeated for a unitary matrix (\ie leading to the maximum possible value for $s$) and for a upper triangular unitary matrix (\ie leading to the minimum possible value for $s$). The value of $h\ =\ 2/3$ used in this work, chosen for consistence to earlier works, leads to a middle point between the two extremes considered and thus provide a good discriminating condition when using the statistics of a random matrix as a null hypothesis.
 \label{fig:heffect}}  }
\end{figure}

\end{document}